\documentclass[12pt]{article} 
\usepackage{amsmath}
\usepackage{amssymb}
\usepackage{amsfonts}
\usepackage{multirow}
\usepackage{natbib}
\usepackage{amsthm}
\usepackage{graphicx}
\usepackage{booktabs}
\usepackage[tight]{subfigure}
\usepackage{caption}
\usepackage{url}
\usepackage{setspace}
\usepackage{mathrsfs}
\usepackage[svgnames,usenames,dvipsnames]{xcolor}
\usepackage{mathtools}
\usepackage{graphicx}
\usepackage[margin=1in]{geometry}
\usepackage{float}
\usepackage{listings}
\usepackage{rotating}
\usepackage{subfigure}
\usepackage{xr}

\newcommand{\be}{\begin{equation}}
\newcommand{\ee}{\end{equation}}
\newcommand{\bea}{\begin{eqnarray}}
\newcommand{\eea}{\end{eqnarray}}
\newcommand{\mbf}[1]{\mathbf{#1}}
\newcommand{\mbs}[1]{\boldsymbol{#1}}

\newcommand{\ba}{\mbf{a}}
\newcommand{\bb}{\mbf{b}}

\newcommand{\bd}{\mbf{d}}

\newcommand{\bv}{\mbf{v}}
\newcommand{\bx}{\mbf{x}}

\newcommand{\bA}{\mbf{A}}

\newcommand{\bB}{\mbf{B}}

\newcommand{\bD}{\mbf{D}}

\newcommand{\bV}{\mbf{V}}
\newcommand{\bX}{\mbf{X}}

\newcommand{\bZ}{\mbf{Z}}
\newcommand{\bW}{\mbf{W}}

\newcommand{\bR}{\mbf{R}}
\newcommand{\bP}{\mbf{P}}
\newcommand{\bU}{\mbf{U}}

\newcommand{\bmu}{\mbs{\mu}}
\newcommand{\bbeta}{\mbs{\beta}}

\newcommand{\blambda}{\mbs{\lambda}}
\newcommand{\bLambda}{\mbs{\Lambda}}

\newcommand{\bet}{{\mbs{\eta}}}

\newcommand{\bI}{\mbf{I}}

\newcommand{\ben}{\begin{equation*}}
\newcommand{\een}{\end{equation*}}
\newcommand{\bean}{\begin{eqnarray*}}
\newcommand{\eean}{\end{eqnarray*}}
\newcommand{\bsm}{\begin{smallmatrix}}
\newcommand{\esm}{\end{smallmatrix}}
\newcommand{\bmat}{\begin{matrix}}
\newcommand{\emat}{\end{matrix}}

\newcommand{\given}{\,|\,}

\newcommand{\bzero}{\mbs{0}}
\newcommand{\bone}{\mbs{1}}
\newcommand{\bell}{\mbs{\ell}}

\newcommand{\bSigma}{\mbs{\Sigma}}

\newcommand\indsim{\stackrel{\mathclap{ind}}{\sim}}
\newcommand\iidsim{\stackrel{\mathclap{iid}}{\sim}}

\begin{document}

   \begin{center}
       \vspace*{1cm}
       \large
	    \textbf{Spatio-temporal areal models to support small area estimation: An application to national-scale forest carbon monitoring}\\
        \normalsize
          \vspace{5mm}
	    Elliot S. Shannon\textsuperscript{1, 2}, Andrew O. Finley\textsuperscript{1, 2}, Paul B. May\textsuperscript{3},\\Grant M. Domke\textsuperscript{4}, Hans-Erik Andersen\textsuperscript{5}, George C. Gaines, III\textsuperscript{6}, Sudipto Banerjee\textsuperscript{7}
	   \\
        \vspace{5mm}
   \end{center}
   {\small 
   \begin{enumerate}
       \item Department of Forestry, Michigan State University, East Lansing, MI, USA.
       \item Department of Statistics and Probability, Michigan State University, East Lansing, MI, USA.
        \item Department of Mathematics, South Dakota School of Mines and Technology, Rapid City, SD, USA.
        \item USDA Forest Service, Northern Research Station, St. Paul, MN, USA. 
        \item USDA Forest Service, Pacific Northwest Research Station, Seattle, WA, USA.
        \item USDA Forest Service, Rocky Mountain Research Station, Missoula, MT, USA.
        \item Department of Biostatistics, University of California, Los Angeles, Los Angeles, CA, USA.
       \end{enumerate}
       }
       \noindent \textbf{Corresponding Author}: Elliot S. Shannon, email: shann125@msu.edu.
         
\section*{Abstract}
\begin{enumerate}
    \item National Forest Inventory (NFI) programs can provide vital information on the status, trend, and change in forest parameters. These programs are being increasingly asked to provide forest parameter estimates for spatial and temporal extents smaller than their current design and accompanying design-based methods can deliver with desired levels of uncertainty. Many NFI designs and estimation methods focus on status and are not well equipped to provide acceptable estimates for trend and change parameters, especially over small spatial domains and/or short time periods. 
    \item Fine-scale space-time indexed estimates are critical to a variety of environmental, ecological, and economic monitoring efforts. Estimates for forest carbon status, trend, and change are of particular importance to international initiatives to track carbon dynamics.  Model-based small area estimation (SAE) methods for NFI and similar ecological monitoring data typically pursue inference on status within small spatial domains, with few demonstrated methods that account for spatio-temporal dependence needed for trend and change estimation. 
    \item We propose a spatio-temporal Bayesian model framework that delivers statistically valid estimates with full uncertainty quantification for status, trend, and change. The framework accommodates a variety of space and time dependency structures, and we detail model configurations for different settings. 
    \item Through analysis of simulated datasets, we compare the relative performance of candidate models and a traditional direct estimator. We then apply candidate models to a large-scale NFI dataset to demonstrate the utility of the proposed framework for providing unique quantification of forest carbon dynamics in the contiguous United States.  We also provide computationally efficient algorithms, software, and data to reproduce our results and for benchmarking.
\end{enumerate}
\section{Introduction}\label{sec:introduction}

Given the ecological and economic importance of forests, national forest inventory (NFI) programs have been implemented to perform large-scale forest monitoring. Data generated by these programs offer a unique and powerful resource for determining the extent, magnitude, and causes of long-term changes in forest health, timber resources, and forest landowner characteristics \citep{Wurtzebach_2019}. Traditionally, NFI programs provide design-based estimates for forest parameters based on measurements taken on a forest inventory plot network \citep{bechtold_patterson_2005, WestfallFIA_2022, tomppo_2010}. Depending on the desired level of estimate precision, such approaches often require repeated costly measurements over a relatively dense inventory plot network; hence, from a cost efficiency standpoint, there is interest in methods that can deliver comparable inference using fewer inventory plots and remeasurements. At the same time, agencies that administer NFIs are experiencing increased demand for estimates within smaller spatial, temporal, and biophysical extents than design-based inference can reasonably deliver \citep{kohl2006sampling, breidenbach2012small, prisley2021}. Developing estimation methods that support inference on small areas---referred to as small area estimation (SAE)---using NFI data is an active area of research, with considerable progress made in recent years \citep{hou2021updating, coulston2021enhancing, schroeder2014improving, lister2020use, finley2024}. Here, ``small area'' does not necessarily refer to small spatial extents alone, rather, to any domain of interest that contains too few observations to deliver accurate direct estimates. SAE methods are numerous and diverse, although most seek to improve inference in small areas by leveraging observations from both within and outside the domain of interest, auxiliary information correlated with the outcome variables, and statistical models describing their relationship.

Though inferences on small area parameters may proceed from the the probabilistic nature of the sample design (e.g., \citealt{breidt2017model, wojcik2022gregory, affleck2023model}), most contemporary SAE methods rely on statistical models for inference \citep{rao2015small}. Both design- and model-based approaches to small area estimation are well developed and compared in the statistical and forestry/ecology literature \citep[see, e.g.,][]{Sarndal1978, Sarndal2003, Gregoire1989, mcroberts2010probability, dumelle2022}. The design-based approach assumes a fixed finite population that is accessible (in principle without error) through a census if all population units were observed. Randomness is incorporated via the selection of population units into a sample according to a randomized sampling design. This is often effective when variability and dependence across population units are adequately captured by the sampling design, which assigns a probability of selection to each sample. However, if units in the population exhibit associations or dependencies that are too complex to be captured by a sampling design or there is a paucity of data, e.g., due  to prohibitive collection costs, then a model-based approach might be preferable \citep[see, e.g., the developments in][for model-based as well as fully Bayesian perspectives to inference for finite populations]{little2004jasa, Ghosh:2012md, banerjee2023finite}. In applications considered here, we assume the population is a realization from a data-generating stochastic process, and hence we pursue model-based SAE.

SAE methods can generally be classified into two groups: unit-level and area-level models. Unit-level models are constructed at the level of population units, which are defined as the minimal units that can be sampled from a population. With respect to most NFI surveys, field plot centers represent population units. Unit-level models typically relate outcome variable measurements on sampled population units to auxiliary data that is available for all population units. Prediction for a small area is achieved by aggregating unit-level predictions within a given areal extent \citep{rao2015small}. In contrast, area-level models are constructed across areal units, where relationships are built between area-specific direct estimates (e.g., generated using design-based estimators applied to samples within each area) and auxiliary data \citep{rao2015small}. Hence, area-level models effectively ``adjust'' direct estimates given auxiliary information. 

For NFI data indexed in space and time, unit-level models often provide additional flexibility for model specification and benefit from the well developed theory and methods for point-referenced spatio-temporal data \citep{banerjee_carlin_gelfand, cressie2011statistics}. Residual spatial and/or temporal dependence in NFI data have been effectively modeled using Gaussian processes (GPs)\citep[see, e.g., ][that have also focused on scaling analysis to massive datasets]{datta_et_al_2016, may_finley_2024, finley2024}. However, if sampling locations are unavailable or data are reported for predefined areas, then area-level models are frequently pursued. Given the often proprietary nature of NFI data, our current work focuses on area-level models, specifically the Fay-Herriot (FH) model \citep{fay_herriot_1979}.

FH models incorporate design-based (i.e., direct) estimates, area-specific auxiliary information, and spatio-temporal dependence among areal units to improve inference, and have recently been applied to forest inventory applications \citep[see, e.g.,][]{VERPLANCK2018287, tomesgen_2021, cao_et_al_2022,Stanke_2022}. Working within a hierarchical Bayesian framework, \cite{VERPLANCK2018287} included conditional autoregressive spatial random effects in a FH model to improve inference for above-ground forest carbon parameters within small areas. This improvement was also shown in \cite{chandra_et_al_2015}, where regression coefficients varied spatially to accommodate nonstationary spatial dependence. \cite{tomesgen_2021} demonstrated a FH model's utility in operational forestry by coupling field and remotely sensed data to improve stand-level estimates. More recently, \cite{may_et_al_2023} developed a Bayesian FH model to estimate forest above-ground biomass density across the contiguous United States (CONUS). Their work identified useful remotely sensed auxiliary predictor variables and advantages to modeling nonstationary relationships using spatially-varying regression coefficients. 

Given that NFI data often cover large spatial extents and temporally dynamic forest and land use systems, posited models should accommodate spatial-temporal dependence. This dependence should likely extend beyond space- and time-varying intercepts to accommodate nonstationary relationships between the outcome and predictor variables \citep[as shown in][]{may_et_al_2023, finley_et_al_2011, datta_et_al_2016, may_finley_2024}. There is a rich literature on spatio-temporal area-level models, much of which originates from public health research \citep[see, e.g.,][]{waller_1997, rushworth_et_al_2014, rushworth_et_al_2017, lee_et_al_2018}. There are, however, few examples of spatio-temporal FH models. \cite{rao_and_yu_1994} extended a FH model to accommodate time-series data, although spatial dependence was not considered. More recently, \cite{MARHUENDA2013308} developed a spatio-temporal FH model for income data in Spain, where the model incorporated a space- and time-varying intercept. 

We intend to provide useful inference for key forest parameters in small areas that cannot be reliably obtained using design-based methods. To this end, we develop a spatio-temporal Bayesian FH model capable of delivering statistically valid estimates with full uncertainty quantification for: 1) annual parameters (e.g., status at a given time); 2) trend (e.g., average change over some time interval); and 3) change (e.g., difference between two points in time). The model and approach to parameter estimation were chosen to accommodate features common in NFI data, including: 1) spatial and temporal dependence across areal units; 2) nonstationary relationships with area-level predictor variables; and 3) missing direct estimates due to the often extremely small spatial and temporal extents. Here, too, we provide computationally efficient parameter estimation algorithms and associated code for implementation. A simulation study is used to assess the proposed model-based estimator, along with its submodels, and compare them with a design-based direct estimator. Finally, the proposed model and submodels are applied to United States Department of Agriculture (USDA) Forest Service NFI data to quantify annual status, trend, and change estimates in county-level carbon density and totals across the CONUS over 14 years.

The remainder of the paper is structured as follows. In Section \ref{sec:data_and_models}, we discuss aspects of NFI data, define the proposed model and submodels, and outline their implementation. The simulation and FIA data analyses are presented in Section \ref{sec:analysis_results}. Results are given in Section \ref{sec:summary} along with a discussion about strengths and weaknesses of the proposed FH model for application in NFI. 

\section{Methods}\label{sec:data_and_models}

\subsection{Direct Estimators}\label{sec:direct}

Here, we consider spatio-temporal NFI survey data that are collected for a design-based inventory system, where individual inventory plots are repeatedly measured within nonoverlapping discrete areal units and time steps. Specifically, let $y_{i,j,t}$ be the measurement for the $i^\text{th}$ sampling unit in areal unit $j$ at time $t$, where $i = 1, \ldots, n_{j,t}$ indexes sampling units, $j = 1, \ldots, J$ indexes areal units, and $t = 1, \ldots, T$ indexes time steps. Hence, $n_{j, t}$ is the total number of sampling units measured in areal unit $j$ at time $t$. Our goal is to learn about a latent parameter of interest, denoted $\mu_{j,t}$, which is the population mean for areal unit $j$ at time $t$. In the SAE setting, the $n_{j,t}$ are too small to produce estimates with desired levels of accuracy, but design-based direct estimates may still be calculated and used to inform the subsequent model-based estimate for $\mu_{j,t}$ presented in Section~\ref{sec:model}. The design-based direct estimate for $\mu_{j,t}$ is calculated as
\begin{equation}\label{direct:mean}
\hat{\mu}_{j,t} = \frac{1}{n_{j,t}}\sum_{i = 1}^{n_{j,t}}y_{i,j,t}.
\end{equation}
where the estimate variance for (\ref{direct:mean}) is 
\begin{equation}\label{direct:var}
\hat{\sigma}_{j,t}^2 =  \frac{1}{n_{j,t}(n_{j,t}-1)}\sum_{i = 1}^{n_{j,t}}(y_{i,j,t} - \hat{\mu}_{j,t})^2.
\end{equation}
 
Often, the areal and/or temporal extent is especially small, and few or no plot measurements are available (i.e., $n_{j,t} \in \{0,1\}$), leading to missing direct estimates $\hat{\mu}_{j,t}$ and/or $\hat{\sigma}^2_{j,t}$. Additionally, when plot measurements in areal unit $j$ at time $t$ are identical, $\hat{\sigma}^2_{j,t}$ will be equal to $0$. In these cases, we still would like the SAE model to produce an estimate for $\mu_{j,t}$. 

\subsection{Model}\label{sec:model}

We hope to learn about $\mu_{j,t}$ from its direct estimate $\hat{\mu}_{j,t}$, its associated variance statistic $\hat{\sigma}_{j,t}^2$, and from information borrowed from direct estimates for adjacent areas proximate in space and time. Further, we might glean information about $\mu_{j,t}$ from relationships between $\hat{\mu}_{j,t}$ and area-level predictors. Given the large spatial and temporal extents of NFI data, we expect the impact of these predictors to be nonstationary over space and/or time. Since we consider only discrete space and time, we employ autoregressive structures to capture spatial and temporal correlation. Spatial correlations are usually modeled through structures such as conditional (CAR) or simultaneous (SAR) autoregressive models \citep{banerjee_carlin_gelfand, ver_hoef_car_sar}. Similarly, temporal correlations are modeled using autoregressive or dynamic structures. Incorporating spatial and/or temporal correlations in these ways could further improve model estimates for $\mu_{j,t}$, especially in cases where direct estimates are imprecise or missing.
 
The model we propose is an extension to the traditional FH model to accommodate spatio-temporal data of the form described above. For areal unit $j$ at time $t$ the model is  
\begin{align}
\hat{\mu}_{j,t} &= \mu_{j,t} + \delta_{j,t},\label{mod:FH_obs}\\ 
\mu_{j,t} &= \beta_0 + \eta_{0,j,t} + \sum_{k = 1}^{p}{x_{k,j,t}\beta_k} + \sum_{k = 1}^{q}{\tilde{x}_{k,j,t}\eta_{k,j,t}} + \epsilon_{j,t}, \label{mod:FH_latent}
\end{align}
where $\delta_{j,t}$ and $\epsilon_{j,t}$ are mutually exclusive error terms distributed as mean zero normal ($N$) distributions with $\delta_{j,t} \indsim N(0, \sigma^2_{j,t})$ and $\epsilon_{j,t} \iidsim N(0, \sigma^2_\epsilon)$. The regression component in \eqref{mod:FH_latent} comprises a customary fixed effects regression component $\beta_0 + \sum_{k = 1}^{p}{x_{k,j,t}\beta_k}$ and a spatio-temporally varying regression component $\eta_{0,j,t} + \sum_{k = 1}^{q}{\tilde{x}_{k,j,t}\eta_{k,j,t}}$. Each $x_{k,j,t}$, for $k=1,\ldots,p$, represents a predictor (or explanatory) variable referenced for each areal unit $j$ and time $t$ that comprises the fixed effects regression and $\tilde{x}_{k,j,t}$ represents one of $q \leq p$ predictor variables included in the spatio-temporally varying component of the model. The restriction $q\leq p$ is not essential, but customary in spatio-temporally varying coefficient models since the $\tilde{x}_{k,j,t}$'s are usually a subset of the $x_{k,j,t}$'s. A Bayesian specification will be completed with prior distributions on all these parameters, where $\eta_{0,j,t}$ and $\eta_{k,j,t}$'s will each be endowed with a spatio-temporal distribution.

As defined in (\ref{direct:mean}), the design-unbiased estimate for $\mu_{j,t}$, i.e., $\hat{\mu}_{j,t}$, is based on the survey design and is assumed to provide information about the true, but latent, mean. Traditionally in FH model applications, the design-based sampling-error variance $\hat{\sigma}^2_{j,t}$ estimate, defined in (\ref{direct:var}), is set as the variance of $\delta_{j,t}$ \citep[see, e.g.,][]{wang_2012, poter_2014, rao2015small}. This, however, can have undesirable inferential consequences as exposed by substituting the expression for $\mu_{j,t}$ in \eqref{mod:FH_latent} into \eqref{mod:FH_obs}. The resulting residual variance is $\mbox{var}(\epsilon_{j,t} + \delta_{j,t}) = \mbox{var}(\epsilon_{j,t}) + \hat{\sigma}_{j,t}^2$, where the latter term is now a fixed constant. The estimate for this residual variance depends on the least squares estimates for $\beta_0$, $\beta_k$'s, $\eta_{0,j,t}$, and $\eta_{k,j,t}$'s that, in turn, depend upon the predictor variables in \eqref{mod:FH_latent}. There appears to be no theoretical assurance that these predictor variables will indeed provide a residual variance estimate that is greater than the fixed estimate $\hat{\sigma}_{j,t}^2$. It seems, then, possible that the estimate for $\mbox{var}(\epsilon_{j,t})$ could be forced to be zero (since it cannot be negative) as a purely numerical consequence of fixing a variance component rather than allowing the data to estimate it.   

Instead, we prefer to incorporate information from the design-based estimate by modeling this variance parameter as an inverse-Gamma ($IG$) random variable $\sigma^2_{j,t} \sim IG\left(\frac{n_{j,t}}{2}, \frac{(n_{j,t}-1)\hat{\sigma}^2_{j,t}}{2}\right)$. Modeling $\sigma^2_{j,t}$ in this way allows us to more clearly obtain potential information from the observed sample size $n_{j,t}$. Specifically, as $n_{j,t}$ increases, the mean of $\sigma^2_{j,t}$ concentrates near $\hat{\sigma}^2_{j,t}$ and its precision increases.  

We now turn to the specifications for spatial and temporal random effects. Following (\ref{mod:FH_latent}), a given specification will have $(q+1)$ random effects vectors, one for the intercept and $q$ for those predictor variables for which we posit there are space- and/or time-varying relationships with the outcome. Here, we develop specifications for one vector of random effects, and then generalize to $(q+1)$ when connecting them to (\ref{mod:FH_latent}). We define a $J\times 1$ vector of spatial random effects as
\begin{equation}\label{mod:space}
    \bet^{s} \sim MVN\left(\bzero, \sigma^{2}_{\eta^s}\,\bR(\rho_{\eta^s})\right),
\end{equation}
where the $s$ superscript on $\bet$ indicates these are the spatial effects following a mean zero multivariate Normal ($MVN$) distribution with covariance matrix $\sigma^{2}_{\eta^s}\,\bR(\rho_{\eta^s})$. Here, and in subsequent notation, bold font indicates a vector or matrix. In this specification, $\sigma^{2}_{\eta^s}$ is the scalar variance, $\rho_{\eta^s}$ is the correlation parameter, and $\bR(\rho_{\eta^s}) = \left(\bD - \rho_{\eta^s} \bW\right)^{-1}$ is the $J \times J$ correlation matrix reflecting a CAR spatial structure, as specified in \cite{banerjee_carlin_gelfand}. Here, $\bW$ is the $J \times J$ binary spatial adjacency matrix with elements $w_{ij} = 1$ if areal units $i$ and $j$ are neighbors and $w_{ij} = 0$ otherwise, with $w_{ii} = 0$. In addition, $\bD$ is a $J \times J$ diagonal matrix with $i^\text{th}$ diagonal element $s_i$, where $s_i$ is the total number of neighbors adjacent to areal unit $i$, with $s_i \geq 1$. While we define the CAR model and subsequent random effects using the covariance matrix, estimation algorithms work with the precision matrix because it yields certain computational advantages detailed in Section \ref{sec:implementation}.

When collecting all $N = JT$ space and time observations, we stack by areal units such that the first $T$ elements correspond to areal unit 1 at time points $1, \ldots, T$, and the last $T$ elements correspond to areal unit $J$ at time points $1, \ldots, T$. We then define an $N\times 1$ vector of area-specific temporal random effects as
\begin{equation}\label{mod:time}
    \bet^{t} \sim MVN\left(\bzero, \sigma^{2}_{\eta^{t}}\,\bI \otimes \bA(\alpha_{\eta^{t}})\right),
\end{equation}
where the $t$ superscript on $\bet$ indicates these are temporal effects, $\sigma^{2}_{\eta^{t}}$ is a scalar variance, $\bI$ is a $J\times J$ identity matrix, $\otimes$ is the Kronecker product operator, and $\bA(\alpha_{\eta^{t}})$ is a $T\times T$ first order autoregressive correlation matrix with temporal correlation parameter $\alpha_{\eta^{t}}$ and $ij^\text{th}$ element equal to $\alpha_{\eta^{t}}^{|i-j|}$. Finally, the $N\times 1$ vector of spatial-temporal random effects is
\begin{equation}\label{mod:mult}
    \bet^{st} \sim MVN\left(\bzero, \sigma^{2}_{\eta^{st}}\bR(\rho_{\eta^{st}})\otimes\bA(\alpha_{\eta^{st}})\right),
\end{equation}
where $\bet^{st}$ indicates these are spatio-temporal effects, $\sigma^2_{\eta^{st}}$ is a scalar variance, and remaining terms have been defined earlier. Hence, the area-specific spatial effect evolves over time.

In subsequent analyses, we consider the following candidate models for $\mu_{j,t}$. 
\begin{align}
    \text{Full model: }&\mu_{j,t} = \beta_0 + \eta^{st}_{0,j,t} + \sum_{k = 1}^{p}{x_{k,j,t}\beta_k} + \sum_{k = 1}^{q}{\tilde{x}_{k,j,t}\eta^{s}_{k,j}} + \epsilon_{j,t}\label{mod:full}\\
    \text{Submodel 1: }&\mu_{j,t} = \beta_0 + \eta^{st}_{0,j,t} + \sum_{k = 1}^{p}{x_{k,j,t}\beta_k} + \epsilon_{j,t}\label{mod:sub1}\\
    \text{Submodel 2: }&\mu_{j,t} = \beta_0 + \eta^{t}_{0,j,t} + \sum_{k = 1}^{p}{x_{k,j,t}\beta_k} + \epsilon_{j,t}\label{mod:sub2}
\end{align}
As described later in Section~\ref{sec:analysis_results}, our current setting displays no evidence of predictor variables $\tilde{x}_{k,j,t}$'s having time-varying impact, hence our full model stops short of specifying spatio-temporal coefficients and considers only space-varying coefficients (SVCs). In other settings, one might consider a $\eta^{st}_{k,j,t}$ associated with some or all $\tilde{x}_{k,j,t}$'s. One might also define a submodel with a space-varying intercept; here, however, because we have repeated measurements and are interested in trend and change within each areal unit, we acknowledge temporal dependence since a space-varying intercept alone would not be adequate.

\subsection{Parameter estimation and inference}\label{sec:parameters}

To complete the Bayesian model specification, we assign prior distributions to the model parameters. For the full model (\ref{mod:full}), the joint posterior distribution for all parameters is
\begin{align}\label{mod:td_mult}
\prod^{J}_{j=1}&\prod^T_{t=1} N\left(\hat{\mu}_{j,t}\given \mu_{j,t},\, \sigma^2_{j,t}\right)\times\prod^{J}_{j=1}\prod^T_{t=1} N\left(\mu_{j,t} \given \beta_0 + \eta^{st}_{0,j,t} + \sum_{k = 1}^{p}{x_{k,j,t}\beta_k} + \sum_{k = 1}^{q}{\tilde{x}_{k,j,t}\eta^{s}_{k,j}},\, \sigma^2_\epsilon\right) \times \nonumber\\
&\prod^{p}_{k = 0}N\left(\beta_k \given \mu_\beta,\, \sigma^2_\beta \right) \times
\prod^{J}_{j=1}\prod^T_{t=1} IG\left(\sigma^2_{j,t} \given \frac{n_{j,t}}{2}, \frac{\left(n_{j,t}-1\right)\hat{\sigma}^2_{j,t}}{2}\right) \times IG\left(\sigma^2_{\epsilon} \given a_\epsilon, b_\epsilon \right) \times \nonumber\\
&MVN\left(\bet^{st}_0 \given \bzero, \sigma^{2}_{\eta^{st}}\bR(\rho_{\eta^{st}})\otimes\bA(\alpha_{\eta^{st}})\right) \times\nonumber\\
& IG\left(\sigma^2_{\eta^{st}_0} \given a_{\eta^{st}_0}, b_{\eta^{st}_0}\right)\times U\left(\rho_{\eta^{st}_0} \given a_\rho, b_\rho\right)\times U\left(\alpha_{\eta^{st}_0} \given a_\alpha, b_\alpha\right)\times\nonumber\\ 
&\prod^{q}_{k=1}MVN\left(\bet^{s}_k \given \bzero, \sigma^{2}_{\eta^s_k}\bR(\rho_{\eta^s_k})\right) \times\prod^{q}_{k=1}IG\left(\sigma^2_{\eta^s_k} \given a_{\eta^s}, b_{\eta^s}\right)\times\prod^{q}_{k=1}U\left(\rho_{\eta^s_k} \given a_\rho, b_\rho\right).
\end{align}
Here, the normal prior distribution mean $\mu_\beta$ and variance $\sigma^2_\beta$ for elements in $\bbeta$ were set at 0 and 10$^5$, respectively. With the exception of the $IG$ prior for $\sigma^2_{j,t}$ described in Section \ref{sec:model}, each scalar variance parameter was assigned an $IG$ prior with a shape $a=2$ and scale $b$ set based on exploratory analysis. With a shape of 2, the $IG$ has an infinite variance and mean centered on the scale parameter. The lower and upper bounds for uniform ($U$) prior distributions on the first order autoregressive parameters $\alpha$ and CAR correlation parameters $\rho$ were set to 0 and 1, respectively.

Parameter inference was based on Markov chain Monte Carlo (MCMC) samples from posterior distributions. With the exception of the correlation parameters $\alpha$ and $\rho$, closed form full conditional distributions are available for all parameters, and hence we sample from the posterior distributions using Gibbs steps. A Metropolis algorithm is used to sample correlation parameters' posterior distributions. Sampling algorithm details are given in Section~\ref{sec:implementation}. As reported in Section~\ref{sec:analysis_results}, posterior inference is based on $M=$ 6,000 post-convergence and thinned samples from three MCMC chains, i.e., 2,000 from each chain. We use convergence diagnostics and thinning rules outlined in \cite{gelman2013}. Point and interval summaries for model parameters presented in Section~\ref{sec:analysis_results} include posterior means, medians, and 95\% credible intervals.

Along with the latent mean $\mu_{j,t}$, we are interested in its temporal trend and change within and among areal units. Importantly, capturing the temporal correlation in the latent mean within areal units is key to generating statistically valid trend and change parameter estimates. An intuitive measure of trend is the linear least squares slope fit to the latent means over time, which is computed for the $j^{\text{th}}$ areal unit as 
\begin{equation}
\theta_j = \frac{\sum^T_{t=1}\left(t - \bar{t}\right)\left(\mu_{j,t} - \bar{\mu}_{j}\right)}{\sum^T_{t = 1}\left(t - \bar{t}\right)^2},
\label{eq:mu_theta}
\end{equation}
where $\bar{t} = \frac{1}{T} \sum_{t = 1}^T t$ and $\bar{\mu}_{j} = \frac{1}{T} \sum_{t = 1}^T \mu_{j,t}$.  Posterior inference about $\theta_j$ is accessible using 
\begin{equation}
\theta^{l}_j = \frac{\sum^T_{t=1}\left(t - \bar{t}\right)\left(\mu^{l}_{j,t} - \bar{\mu}^{l}_{j}\right)}{\sum^T_{t = 1}\left(t - \bar{t}\right)^2},
\end{equation}
where $l$ indexes posterior samples $l = 1,2,\ldots,M$. In this way, one sample from $\mu_{j,t}$'s posterior distribution yields one sample from $\theta_j$'s distribution. Given these posterior samples, we can estimate the trend's direction and strength, and perform hypothesis tests, whereby strong support for a non-zero linear trend occurs when the 95\% credible interval for $\theta_j$'s posterior distribution does not include zero.

In the same manner, we obtain estimates for change in the latent mean between two time points $t_1$ and $t_2$ 
within areal unit $j$ as
\begin{equation}
    \Delta_j = \mu_{j,t_2} - \mu_{j,t_1}.
\label{eq:mu_delta}
\end{equation}
Then, we obtain posterior samples $\Delta^{l}_j = \mu^{l}_{j,t_2} - \mu^{l}_{j,t_1}$ for each $l = 1,2,\ldots,M$.

Given an areal unit's density estimate and area, an estimate for the unit's total is generated. Specifically, for $\mu_{j,t}$ expressed on a per unit area basis, with $A_j$ being the known area of the $j^{th}$ areal unit, we calculate $\Omega_{j, t} = A_j\mu_{j,t}$ as our estimate for the total in areal unit $j$ at time $t$. Similarly, if trend and change parameters are expressed as densities per unit area, then their totals are $A_j\theta_{j}$ and $A_j\Delta_{j}$, respectively. As described above, we can collect samples from the posterior distributions of these total parameters. 

Thus far we have considered estimates for individual areal units. However, density, total, trend, and change estimates are easily generated for an arbitrary aggregate of areal units, denoted as $\mathcal{J} \subseteq \{1, \ldots, J \}$. Samples from the posterior distribution of an aggregate's total, $\Omega_{\mathcal{J}, t}$ are obtained simply as 
\begin{equation}
    \Omega^{l}_{\mathcal{J}, t} = \sum_{j \in \mathcal{J}} A_j\mu^{l}_{j,t}.
\label{eq:omega_aggregate}
\end{equation} 
Given (\ref{eq:omega_aggregate}), one can sample from the aggregate's density $\mu_{\mathcal{J}, t}$ posterior distribution using
\begin{equation}
    \mu^{l}_{\mathcal{J}, t} = \frac{\Omega^{l}_{\mathcal{J}, t}} {A_\mathcal{J}},
\label{eq:mu_aggregate}
\end{equation}
where $A_\mathcal{J} = \sum_{j \in \mathcal{J}} A_j$.

Further, aggregate trend $\theta_{\mathcal{J}}$ and change $\Delta_{\mathcal{J}}$ parameters are sampled via
\begin{equation}
    \theta^{l}_{\mathcal{J}} = \frac{\sum^T_{t=1}\left(t - \bar{t}\right)\left(\mu^{l}_{\mathcal{J},t} - \bar{\mu}^{l}_{\mathcal{J}}\right)}{\sum^T_{t = 1}\left(t - \bar{t}\right)^2},
\label{eq:theta_aggregate}
\end{equation}
with $\bar{\mu}^{l}_{\mathcal{J}} = \frac{1}{T} \sum_{t = 1}^T \mu^{l}_{\mathcal{J},t}$, and 
\begin{equation}
    \Delta^{l}_{\mathcal{J}} = \mu^{l}_{\mathcal{J},t_2} - \mu^{l}_{\mathcal{J},t_1},
\label{eq:delta_aggregate}
\end{equation}
respectively. Totals based on trend and change parameters are $A_\mathcal{J}\theta_{\mathcal{J}}$ and $A_\mathcal{J}\Delta_{\mathcal{J}}$, with samples from the respective posterior distributions collected one-for-one using samples from \eqref{eq:theta_aggregate} and \eqref{eq:delta_aggregate}.

\subsection{Model comparison}\label{sec:waic}

In subsequent analyses, we compare candidate models using the widely applicable information criterion \citep[WAIC;][]{Watanabe2010} as implemented in the \texttt{loo} R package \citep{loo2024}. A model's estimate for $\text{WAIC} = -2\widehat{\text{elpd}}_{\text{WAIC}}$ and $\widehat{\text{elpd}}_{\text{WAIC}} = \text{lpd} - \widehat{p}_{\text{WAIC}}$, where lpd is the log pointwise predictive density and $\widehat{p}_{\text{WAIC}}$ is the estimated effective number of parameters. Models with lower WAIC should have better fit to the observed data and yield better out-of-sample prediction. For model comparison, we look to the difference in models' $\widehat{\text{elpd}}_{\text{WAIC}}$ and this difference's standard error estimate, labeled $\widehat{\text{elpd}}_{\text{diff}}$ and $\widehat{\tau}_{\text{diff}}$, respectively, see \cite{Vehtari2017} for details.

\subsection{Implementation and computing considerations}\label{sec:implementation}

Gibbs and Metropolis algorithms, used to sample random effects and their process parameters, require increasingly expensive matrix decompositions as $J$ and/or $T$ become large. However, by working with the CAR model's precision matrix, and not the covariance matrix, we are able to avoid the inversion of the covariance matrix in the Gaussian density. Furthermore, the full conditional distributions for the spatial effects are all closed-form Gaussian and the spectral decomposition of the precision matrix lends itself to additional computational gains. MCMC samplers are written in \texttt{R} \citep{R} using sparse matrix routines from the \texttt{Matrix} package \citep{matrix2023}. In cases where direct estimates $\hat{\mu}_{j,t}$ and/or $\hat{\sigma}^2_{j,t}$ are missing (see Section \ref{sec:direct}), the corresponding $\mu_{j,t}$ sample is updated from its posterior predictive distribution (\ref{mod:FH_latent}).  MCMC sampler details and computing considerations are provided  in Section~\ref{sec:sampler_computing}.

All results presented are reproducible using the MCMC sampler code, data, and workflow scripts provided as Supporting Information Section~\ref{sec:supporting_code}. 

\section{Applications}\label{sec:analysis_results}

The United Nations Framework Convention on Climate Change (UNFCCC) requires parties to submit annual reports with estimates of economy-wide greenhouse gas (GHG) emissions and removals \citep{UNFCCC}. Guidelines for reporting in this agreement and others like it (e.g., 2015 Paris Agreement) have been established by the Intergovernmental Panel on Climate Change \citep{IPCC}. Core among the guidelines are that GHG inventories include ``neither over- nor underestimates so far as can be judged, and in which uncertainties are reduced as far as practicable.'' In the United States GHG inventory, the NFI is relied on to produce accurate annual estimates for change in forest parameters such as carbon to meet reporting requirements \citep{Domke_2023}. However, countries, including the United States (US), are increasingly seeking to improve the accuracy and precision of inventory estimates at finer spatial and temporal scales to meet inventory requirements and other stakeholder needs. However, the relatively coarse NFI spatio-temporal sampling intensities have limited advances in estimating interannual variability in carbon stocks and stock changes on forest land. Given these challenges, the novelty and importance of estimating temporal components in NFI data for small areas is amplified, and provides further motivation for our proposed model and data application. 

We present two applications that make use of NFI data collected and developed by the Forest Inventory and Analysis (FIA) program of the USDA Forest Service. An overview of these data are presented in Section~\ref{sec:fia}. The first application considers a simulated population built to mimic FIA data characteristics and used to assess design- and model-based estimator qualities. The second application considers actual FIA annual county-level estimates and explores trend in forest carbon density and change over a 14 year study period across the CONUS.

\subsection{FIA NFI data}\label{sec:fia}

The FIA program measures and monitors more than 300 thousand forest inventory plots in the CONUS \citep{bechtold_patterson_2005, WestfallFIA_2022}. Each year, remeasurements are made at approximately 1/5  or 1/7 of all inventory plots in the eastern US and 1/10 of inventory plots in the west, leading to remeasurement intervals (i.e., inventory cycle lengths) ranging from 5 to 10 years. These data are used to generate design-based estimates for important forest parameters, such as forest carbon densities and totals. Carbon content is estimated using measured tree characteristics and national-scale carbon allometric equations based on a comprehensive database of tree species, characteristics, and spatial patterns, for which species-specific carbon fractions are then applied \citep{westfall_2023}. Here, individual tree above-ground carbon estimates are aggregated at the plot-level and then expanded to a per hectare basis to express carbon density (Mg/ha). Following notation in Section~\ref{sec:direct}, we let $y_{i,j,t}$ represent the estimated carbon density (Mg/ha) for FIA plot $i$ in county $j$ and year $t$. 

\begin{figure}[ht!]
\centering
\includegraphics[trim={0 1cm 0 1cm},clip,width=10cm]{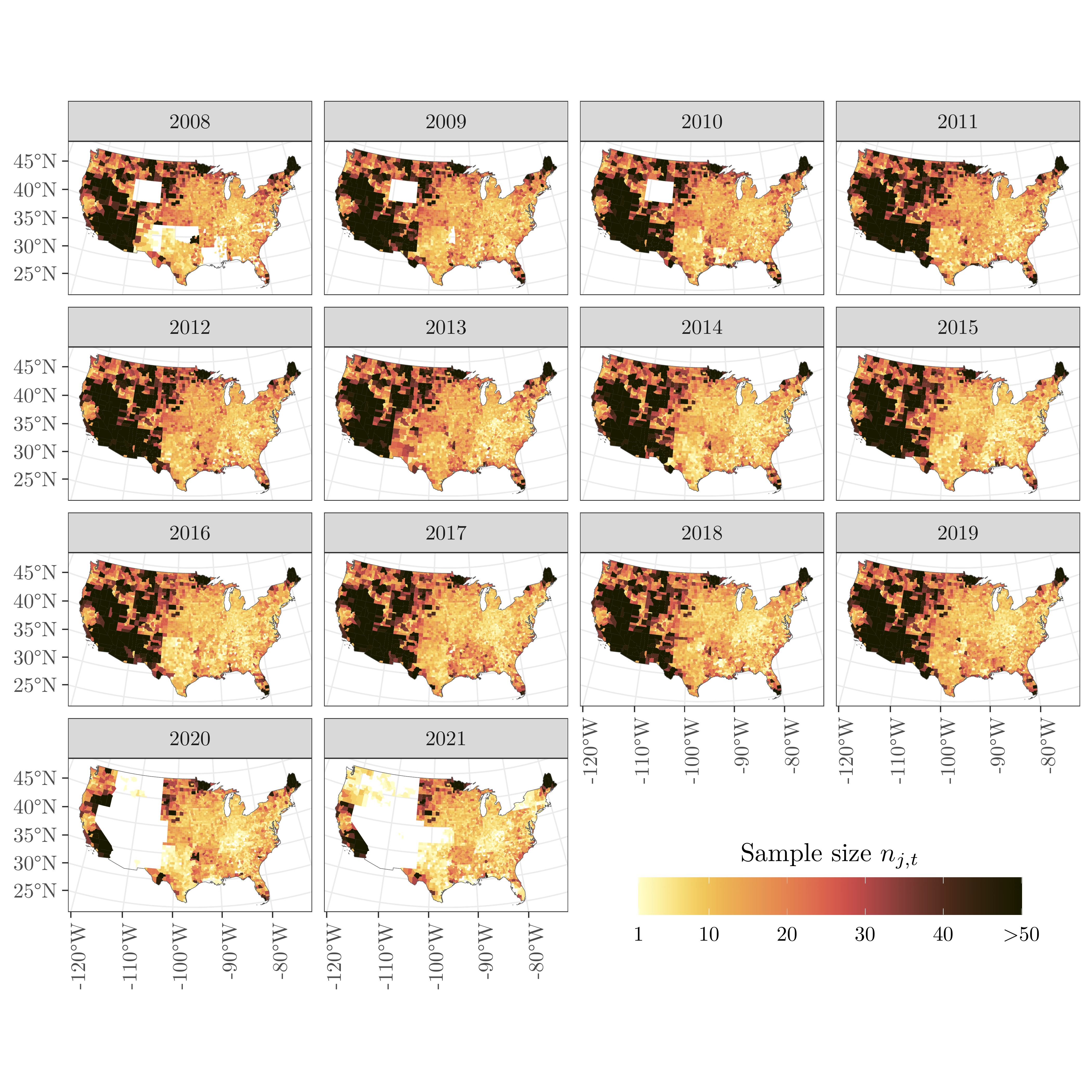}
\caption{Number of observed FIA plots within each county and year. Transparent counties have zero observed FIA plots.}\label{fig:sample_size}
\end{figure}

Here, we consider $J$ = 3,108 counties that comprise the CONUS and $T$ = 14 years of plot measurements from 2008 to 2021. A county-level map of FIA plot sample size (i.e., $n_{j,t}$) is shown in Figure~\ref{fig:sample_size}. Only the base intensity plot network, excluding additional intensification plots collected by some states and agencies, is used in our analysis. Of the $N$=43,512 county and time observations, there are 10,627 for which direct estimates are missing. This missingness arises from the absence of measured FIA plots (number of occurrences 1,860), the absence of sampling error variance due to $n_{j,t} = 1$ (820), or the sampling error variance equaling 0 in areal units with only carbon density measurements of zero (7,947). Figure~\ref{fig:missing} summarizes missingness by county, much of which is concentrated in western and midwestern counties due to lack of forest, survey implementation, and reporting.

\begin{figure}[ht!]
\centering
\includegraphics[trim={0 1cm 0 1cm},clip,width=10cm]{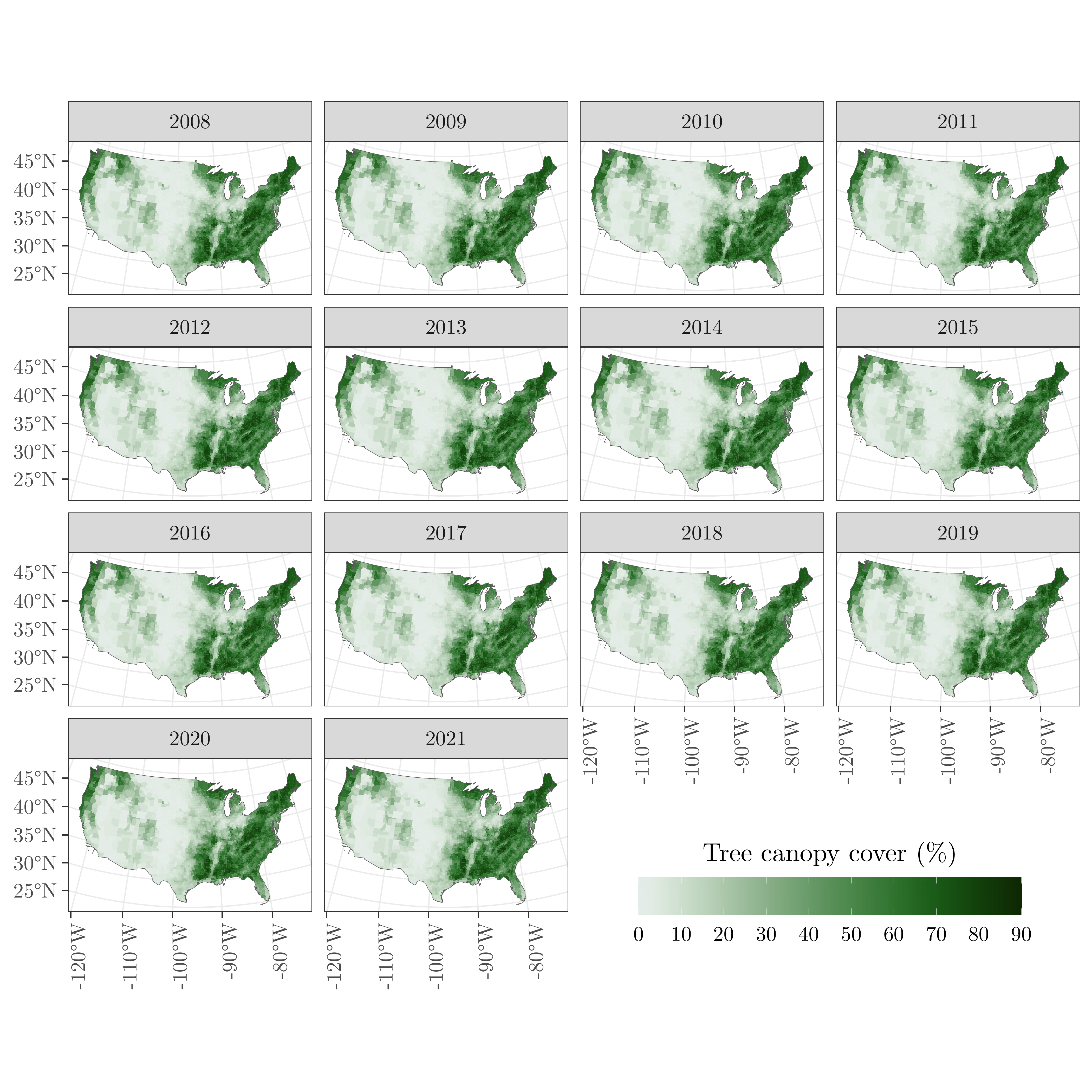}
\caption{Annual National Land Cover Database percent tree canopy cover averaged within each county.}\label{fig:TCC}
\end{figure}

Percent tree canopy cover (TCC) data for the CONUS are produced by the USDA Forest Service as part of the National Land Cover Database (NLCD) \citep{tcc_methods_2023}. The TCC data are derived from multispectral remote sensing data and have been released as annual maps for years 2008 to 2021 at a 30-by-30 (m) pixel spatial resolution for the CONUS. These data represent fractional pixel-level tree canopy cover expressed as a percentage, which we average for each county and year to produce annual county-level mean TCC (\%) shown in Figure~\ref{fig:TCC}. For use in subsequent models, this TCC variable was centered and scaled to have mean zero and variance one, and denoted as $x_{TCC}$ where the $TCC$ subscript takes the place of subscript $k$ in models (\ref{mod:full}-\ref{mod:sub2}). 

For the analysis of FIA data presented in Section~\ref{sec:real}, interpretation of forest carbon density estimates comes with a caveat and limitation. The FIA program defines forest land as land which has at least 10\% canopy cover of trees of any size, or had at least 10\% canopy cover of trees in the past, based on the presence of stumps, snags, or other evidence, and that will be naturally or artificially regenerated. FIA records zero carbon density for non-forest plots, although they could have up to 10\% tree canopy cover. Hence, our direct estimate (computed using both forest and non-forest plots) will be negatively biased. Although this bias might be small, difficulty arises when we wish to attribute change in carbon density to change in forest structure and land use---carbon change results from change in forest land area and/or change in carbon density within forest land (e.g., fewer trees). As articulated by \cite{Wiener2021} and \cite{Knott2023}, this is a known caveat and limitation when using FIA data for SAE, but also an opportunity for FIA to consider expanding data collection protocols to better support SAE and other model-based estimation efforts.  

\subsection{Simulation study overview}\label{sec:simulationOverview}
The simulation study is presented in the Supporting Information Section~\ref{sec:simulation}. Here, a single population is generated using fixed and known values for parameters, then estimates for $\mu_{j,t}$, $\theta_j$, and $\Delta_j$ are generated from each of a large number of independent samples taken from the population. The four estimators considered are the direct (\ref{direct:mean}), Full model (\ref{mod:full}), Submodel 1 (\ref{mod:sub1}), and Submodel 2 (\ref{mod:sub2}). Estimates were compared with the true population parameter values using a set of measures that assess the estimators' bias, accuracy, and precision. The population was simulated to mimic qualities of the observed FIA annual county-level carbon density (Mg/ha) data, with both spatial and temporal dependence. Simulation study results show that: 1) all estimators are biased when the sample size is small; 2) candidate models yield improved accuracy and uncertainty quantification over the direct estimator; 3) models' IG prior on $\sigma^2_{j, t}$ behaves as expected, weighting the direct estimates according to samples size; 4) WAIC is useful for selecting the candidate model that most closely resembles the data generating model.

\subsection{Analysis of FIA NFI data} \label{sec:real}
In this section, we analyze the FIA carbon data described in Section~\ref{sec:fia}. Similar to the simulated data analysis, we consider four estimators for the parameter of interest, $\mu_{j,t}$, i.e., direct (\ref{direct:mean}), Full model (\ref{mod:full}), Submodel 1 (\ref{mod:sub1}), and Submodel 2 (\ref{mod:sub2}), using direct estimates and $x_{TCC}$ predictor variable values. Given estimates for $\mu_{j,t}$, we also estimate derived parameters $\theta_j$ and $\Delta_j$.

As described in Section~\ref{sec:introduction}, we anticipated the NFI data to exhibit strong temporal and spatial dependence in the latent mean within and among areal units. This dependence is reflected in the process parameter estimates in Table~\ref{tab:real_ests} where temporal and spatial process parameters associated with the model intercept show strong correlation (i.e., $\alpha_{\eta^{t}_0}$, $\alpha_{\eta^{st}_0}$, and $\rho_{\eta^{st}_0}$ are close to 1) and relatively large variances (i.e., $\sigma^2_{\eta^{t}_0}$ and $\sigma^2_{\eta^{st}_0}$). For the Full model, the spatial process parameters associated with $x_{TCC}$, i.e., $\rho_{\eta^s_{TCC}}$ and $\sigma^2_{\eta^{s}_{TCC}}$, have a strong correlation and relatively large variance. 

Considering model fit using WAIC given in Table~\ref{tab:real_ests},  Submodel 1 has the lowest WAIC, followed by the Full model and then Submodel 2. When testing differences between Submodel 1 and the other candidate models, we see no difference between Submodel 1 and the Full model (i.e., $-15\pm1.96 \times 22.2$ includes zero) and Submodel 1 is substantially better than Submodel 2 (i.e., $-642.5\pm1.96 \times 51.9$ excludes zero). 

\begin{table}[ht!]
\begin{center}
\begin{tabular}{lccc}
\toprule
&\multicolumn{3}{c}{Candidate models} \\
\cmidrule(lr){2-4} 
Parameter & Submodel 2 & Submodel 1 & Full model\\
\midrule
$\beta_0$ &18.15 (17.92, 18.38)&18.39 (18.00, 18.79)&13.83 (12.48, 15.13)\\
$\beta_{TCC}$ &15.27 (15.04, 15.53)&18.43 (18.16, 18.83)&16.08 (15.05, 17.18)\\\
$\alpha_{\eta^{t}_0}$ &0.9982 (0.9979, 0.9985)&-&-\\
$\sigma^2_{\eta^{t}_0}$ &75.73 (71.61, 80.29)&-&-\\
$\rho_{\eta^{st}_0}$  &-&0.9995 (0.9990, 0.9998)&0.9996 (0.9993, 0.9998)\\
$\alpha_{\eta^{st}_0}$  &-&0.9966 (0.9961, 0.9972)&0.9969 (0.9963, 0.9975)\\
$\sigma^2_{\eta^{st}_0}$ &-&140.37 (130.31, 150.30)&86.81 (79.14, 94.11)\\
$\rho_{\eta^s_{TCC}}$  &-&-&0.9999 (0.9995, 1.0000)\\
$\sigma^2_{\eta^{s}_{TCC}}$ &-&-&34.71 (28.55,  41.29)\\
$\sigma^2_{\epsilon}$  &0.70 (0.61, 0.79)&0.59 (0.54, 0.66)&0.56 (0.51, 0.63)\\
\midrule
$\widehat{\text{elpd}}_{\text{WAIC}}$ &-113123.6&-112481.2&-112496.2\\
$\widehat{p}_{\text{WAIC}}$ &7078.1&6622.7&6461.9\\
WAIC &226247.3&224962.4&224992.4\\
$\widehat{\text{elpd}}_{\text{diff}}$ &-642.5 (51.9)& 0 (0)& -15.0 (22.2)\\
\bottomrule
\end{tabular}
\caption{Parameter estimates for candidate models fit to FIA data. Estimates are posterior medians with 95\% credible intervals given in parentheses. Estimates for WAIC and associated statistics are given in the last several rows. The last row holds the estimated $\widehat{\text{elpd}}_{\text{WAIC}}$ difference with the ``best'' fitting Submodel 1 and associated standard error $\widehat{\tau}_{\text{diff}}$ in parentheses.} \label{tab:real_ests}
\end{center}
\end{table} 

As seen in the simulation study presented in Section~\ref{sec:simulation}, we again observe varied effects of the prior for $\sigma^2_{j,t}$ on its posterior (Figure~\ref{fig:real_sigmaSq_prior}(a)). This variation reflects increased information about the direct estimate for larger sample sizes, and greater discrepancy between prior and posterior for smaller sample sizes. Similarly, when the sample size is large, posterior estimates for $\mu_{j,t}$ are more similar to the direct estimate $\hat{\mu}_{j,t}$, and when the sample size is small, the posterior estimate is informed by other sources including neighboring direct estimates, predictor variables, and structured random effects (Figure~\ref{fig:real_sigmaSq_prior}(b)). Said differently, when the sample size is small, we express our uncertainty in the direct estimate through the prior, and after observing evidence (i.e., data), our posterior estimate better reflects the information available in the data, which is likely different from the prior. This balance highlights the importance of incorporating prior information when it is available, while acknowledging uncertainty when it is not. 

\begin{figure}[ht!]
\centering
\includegraphics[trim={0 3cm 0 0cm},clip,width=12cm]{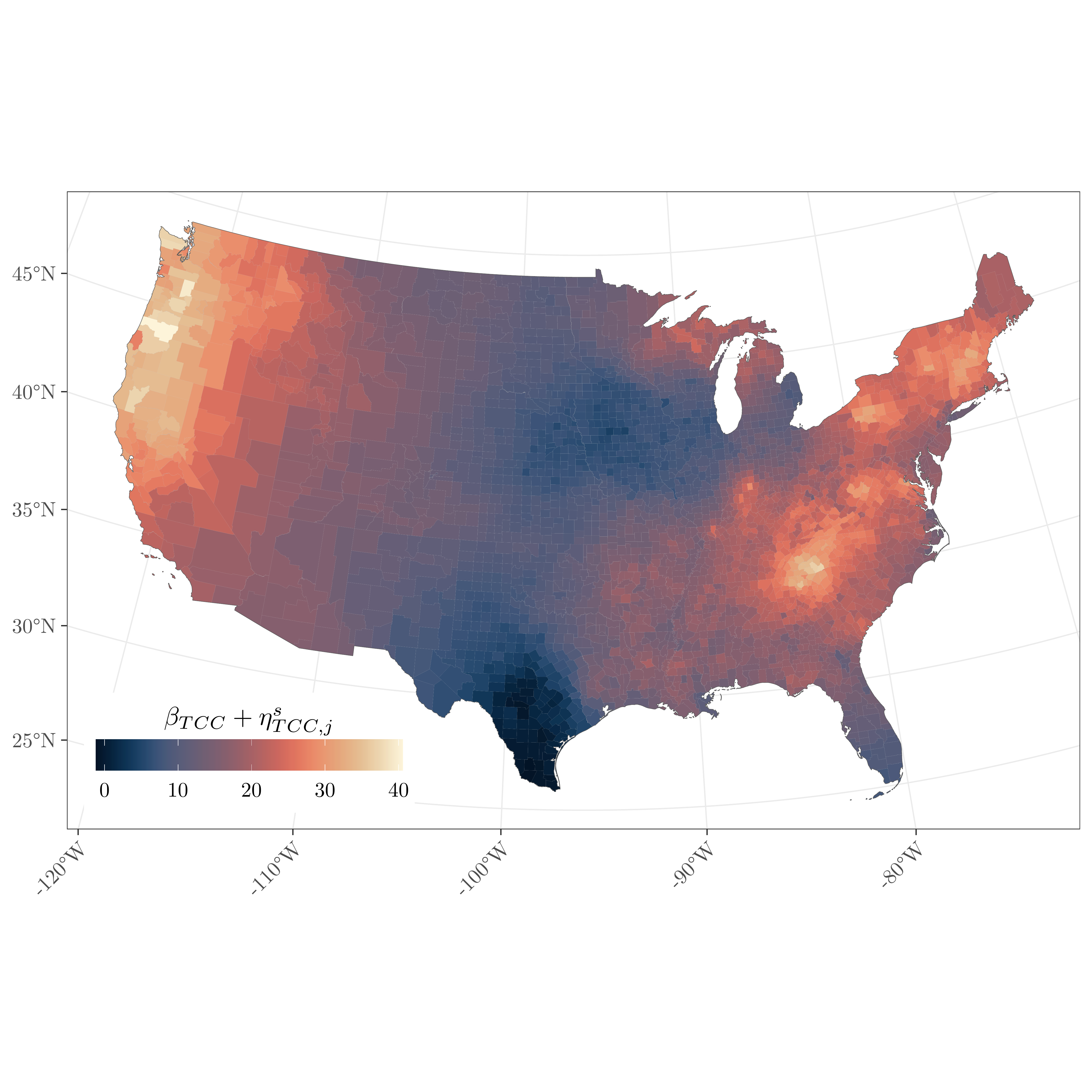}
\caption{Posterior median of the Full model's space-varying coefficient $\beta_{TCC}+\eta^s_{TCC,j}$ fit to the FIA data.}\label{fig:full_tcc_svc}
\end{figure}

Also discussed in Section~\ref{sec:introduction}, given the NFI's large spatial extent that covers many different forest types, environments, and disturbance regimes, we anticipate a nonstationary relationship between tree canopy cover (as captured through $x_{TCC}$) and carbon density. In other words, we expect the relationship between percent TCC and carbon density to vary spatially over the CONUS. In addition, estimates for the Full model's $\rho_{\eta^s_{TCC}}$ and $\sigma^2_{\eta^{s}_{TCC}}$ in Table~\ref{tab:real_ests} do suggest long-range spatial dependence and substantial variability of $\bet^s_{TCC}$, reflecting the mixture of underlying environmental and ecological processes that drive spatial variation in forest carbon density. The Full model's SVC $\beta_{TCC}+\bet^s_{TCC}$, shown in Figure~\ref{fig:full_tcc_svc}, is particularly useful for identifying $x_{TCC}$'s nonstationary impact on carbon density, whereby an increment change in $x_{TCC}$ in northern California, the Pacific Northwest, the southern Appalachian Mountains, and southern New England (i.e., the brighter color regions) is associated with a greater change in carbon density than regions with darker colors. For example, per unit area, a county with 50\% forest canopy cover in the Pacific Northwest, Appalachian Mountains, or New England will have more carbon than a county with 50\% forest canopy cover in the midwest, Texas, or southern Florida.

Given the clear nonstationarity seen in Figure~\ref{fig:full_tcc_svc}, all subsequent results presented are based on the Full model. Figures~\ref{fig:real_carbon_mean} and \ref{fig:real_carbon_sd} map $\mu_{j,t}$'s posterior distribution mean and standard deviation, respectively. For counties that have complete missingness (i.e., no direct estimates in the study period), as depicted in Figure~\ref{fig:missing}, estimates come from the posterior predictive distribution of $\mu_{j,t}$ and will typically have a greater uncertainty than those borrowing information from the given county's direct estimates. This greater uncertainty is most apparent in central and some north-central counties when comparing Figures~\ref{fig:real_carbon_sd} and \ref{fig:missing}. If increasing sample size is not an option, then reducing $\sigma^2_{\epsilon}$, via additional predictor variables or random effects, will reduce the posterior predictive distribution's dispersion.

\begin{figure}[ht!]
\centering
\includegraphics[trim={0 1cm 0 1cm},clip,width=12cm]{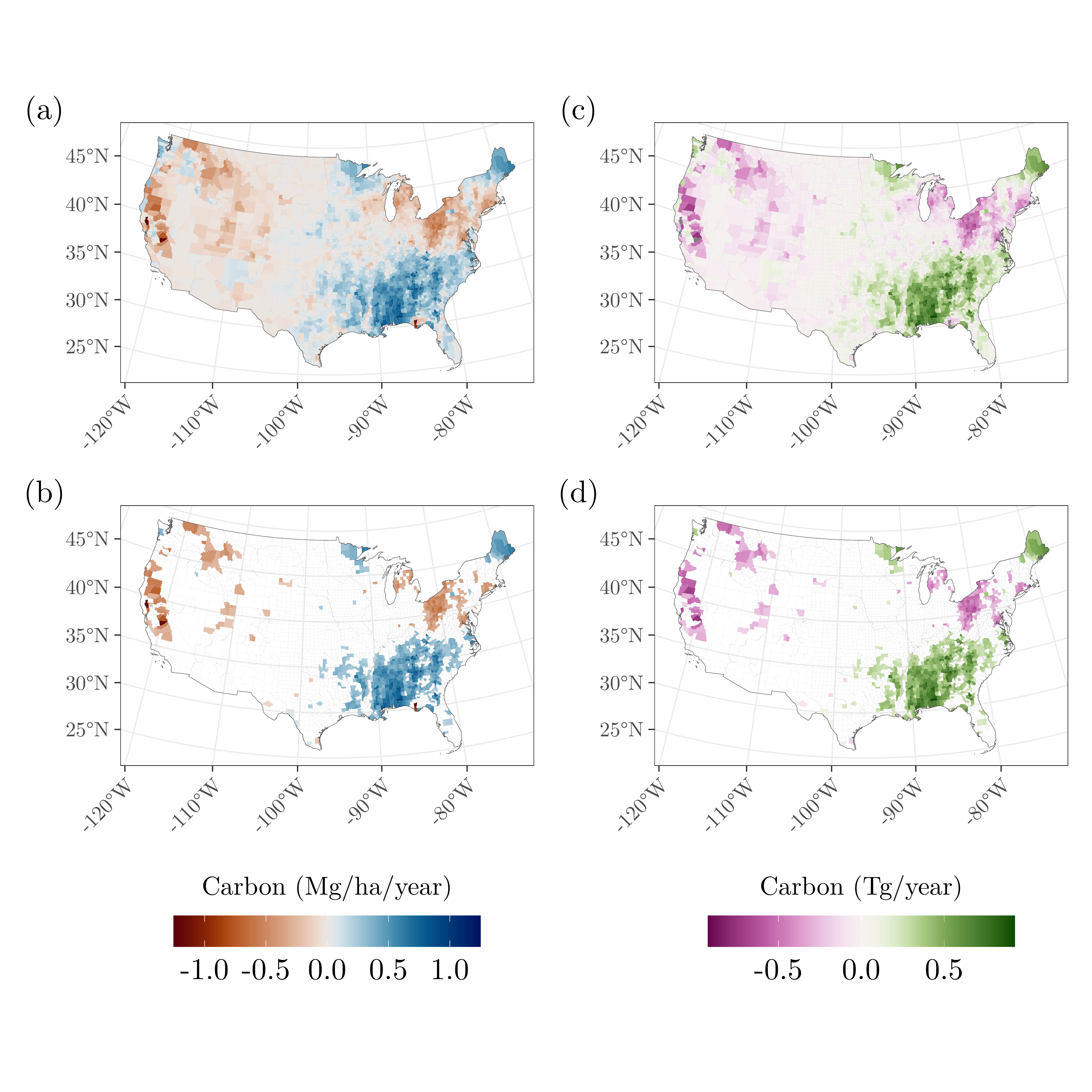}
\caption{Estimates from the Full model. (a) Estimated linear trend in forest carbon density $\theta_j$ (Mg/ha/year). Values are each county's posterior distribution median. (b) Counties from (a) that have posterior distribution 95\% credible intervals that exclude zero. (c) Estimated linear trend in forest carbon total $A_j\theta_j$ (Tg/year), where $A_j$ is the $j^{\text{th}}$ county's area in hectares. (d) Counties from (c) that have posterior distribution 95\% credible intervals that exclude zero. The 10 counties with the largest and smallest values in (b) and (d) are given in Tables~\ref{tab:real_largest_theta} and \ref{tab:real_largest_theta_carbon}, respectively.}\label{fig:real_theta}
\end{figure}

Carbon density trend estimates for $\theta_j$ are shown in Figure~\ref{fig:real_theta}(a). Figure~\ref{fig:real_theta}(b) shows only those county values in Figure~\ref{fig:real_theta}(a) that have a posterior distribution 95\% credible interval that does not include zero. Here, several expected patterns in carbon density increase and decrease emerge. As detailed in \cite{Harris2016}, \cite{Hoover2021}, and similar studies on forest biomass/carbon change in the CONUS, factors including fire, insect, disease, harvest, and changing forest age and land use drive the observed patterns of carbon loss in western counties and the Ohio Valley. In contrast, the increase in carbon density in the southeast and Maine is primarily due to increases in industrial forest operations, changes in forest age and structure, and increased forest productivity \citep{Hoover2021, Hogan_2024}.

\begin{figure}[ht!]
\centering
\includegraphics[trim={0 0cm 0 0cm},clip,width=12cm]{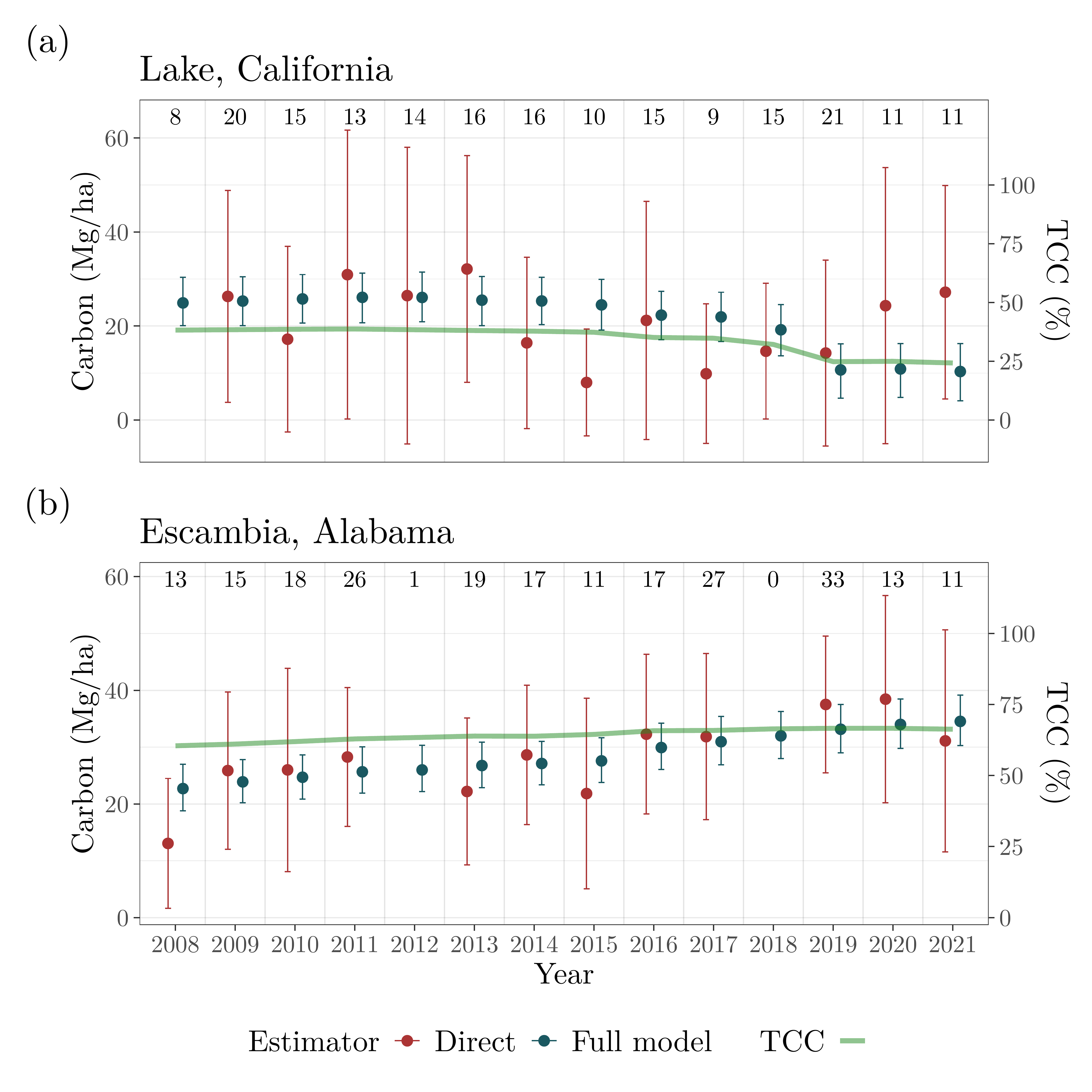}
\caption{Direct and model estimates for $\mu_{j,t}$ along with TCC values for counties with the largest negative and positive $\theta_j$ estimates given in Table~\ref{tab:real_largest_theta}. Sample size $n_{j,t}$ for each county and year is given across the top of each subpanel. Estimate means and medians are shown as points with 95\% confidence and credible
interval bars for direct and model estimates, respectively. When $n_{j,t} \leq 2$ or all observations are zero, the direct estimate is not available, e.g., Escambia, Alabama in 2012 and Lake, California in 2008, respectively.}\label{fig:real_ex_county_theta_pos_neg}
\end{figure}

Figure~\ref{fig:real_ex_county_theta_pos_neg} shows the direct and model estimates for $\mu_{j,t}$ for the two counties with the largest negative and positive $\theta_j$ estimates in Figure~\ref{fig:real_theta}(b). A list of counties with the 10 largest negative and positive $\theta_j$ estimates in Figure~\ref{fig:real_theta}(b) are given in Table~\ref{tab:real_largest_theta}. As recorded in Table~\ref{tab:real_largest_theta}, Lake, California has the largest negative $\theta_j$ estimate among all counties in the CONUS. Over the study period, Lake County's estimate for $\theta_j$ was -1.25 (-1.62, -0.88) (Mg/ha/year). This negative trend is evident in Lake County's $\mu_{j,t}$ estimates shown in Figure~\ref{fig:real_ex_county_theta_pos_neg}(a). Much of Lake County's carbon loss occurred in 2018 as a result of the River, Ranch, and Pawnee Fires that burned an estimated 48,400 (ha) \citep{Tyukavina2022}. The effect of these 2018 fires is clearly seen in the model estimates and TCC values in Figure~\ref{fig:real_ex_county_theta_pos_neg}(a). Given the relatively small sample sizes, it is not surprising that direct estimates do not capture the impact of fire on forest carbon. With lack of direct estimate information, the model learns from TCC, which clearly shows a loss of canopy cover starting in 2018 (Figure~\ref{fig:real_ex_county_theta_pos_neg}(a)). Escambia Alabama has the largest positive $\theta_j$ estimate among all counties in the CONUS. Over the study period, Escambia County's estimate for $\theta_j$ was 0.92 (0.67, 1.20) (Mg/ha/year). This positive trend is seen in Figure~\ref{fig:real_ex_county_theta_pos_neg}(b) and reflects carbon accumulation in the county's intensively managed softwood plantations. Figures comparable to Figure~\ref{fig:real_ex_county_theta_pos_neg} for all counties are provided in Supporting Information Section~\ref{sec:county_supporting_figures}.

Given the county area and our estimate for carbon density, an estimate of total carbon trend is $A_j\theta_j$, where $A_j$ is the area of the $j^\text{th}$ county in hectares. These total county carbon estimates are shown in Figures~\ref{fig:real_theta}(c,d), with the 10 largest negative and positive county estimates given in Table~\ref{tab:real_largest_theta_carbon}. This table shows that, once scaled by county area, large-area counties in California and Maine dominate negative and positive carbon trends, respectively. Moreover, we observe clear regional patterns, with the most negative carbon trends occurring in the west and the most positive in the northeast and southeast.

Following the simulated data analysis, we calculated estimates for $\Delta_j$ and $A_j\Delta_j$ to investigate carbon change over the study period. These change patterns were similar to those for $\theta_j$ and are presented in Figure~\ref{fig:real_delta}. 

\begin{figure}[ht!]
  \centering
  \includegraphics[trim={0 0cm 0 0cm},clip,width=12cm]{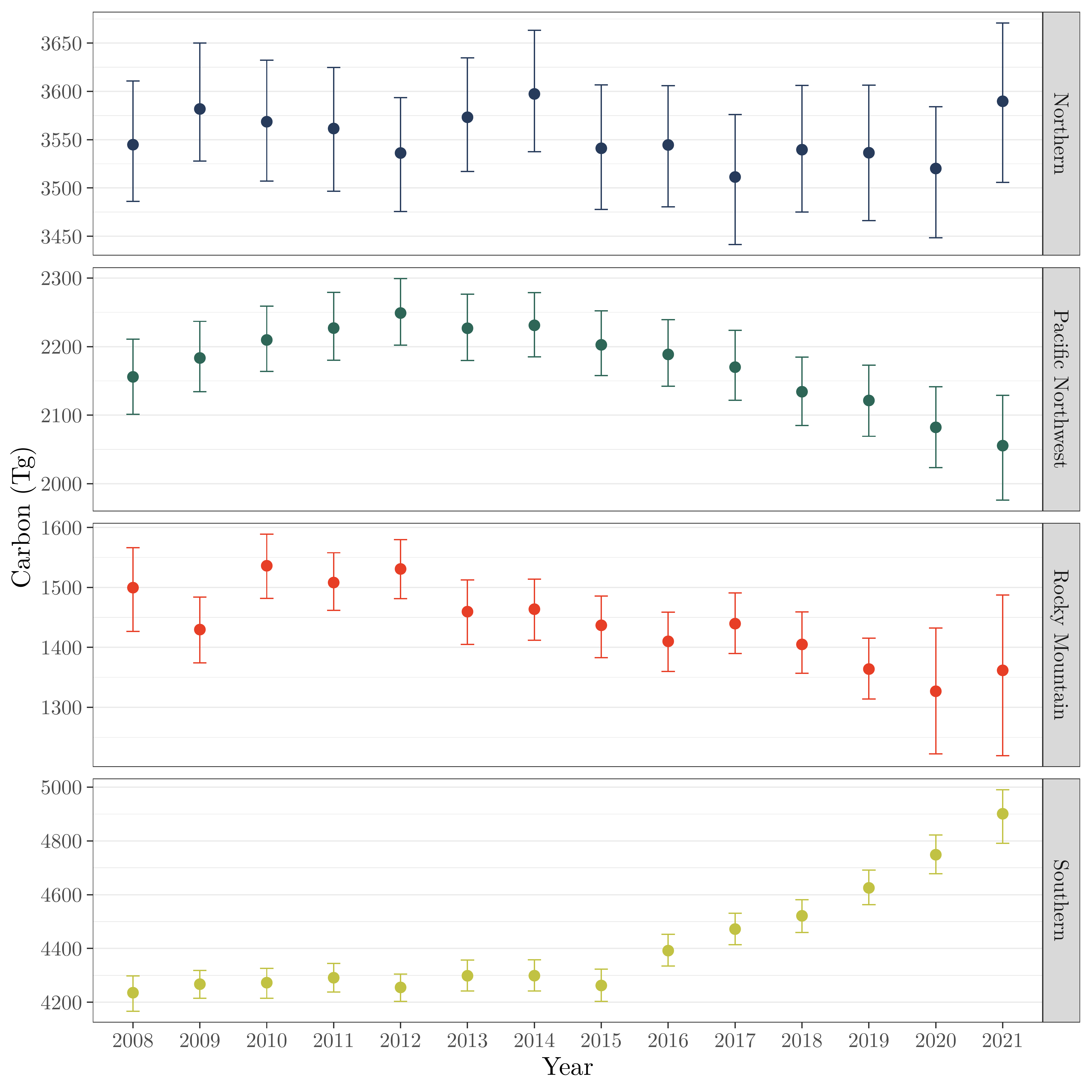}
  \caption{Full model estimates for annual carbon $\Omega_{\mathcal{J},t}$ within FIA regions shown in Figure~\ref{fig:fia_regions}. Point and bars are the median and 95\% credible interval for posterior distributions summarized over region and year following \eqref{eq:omega_aggregate}.}\label{fig:fia_regions_annual_totals}
\end{figure}

For administrative and reporting purposes, FIA partitions the CONUS into four regions shown in Figure~\ref{fig:fia_regions}: Northern; Pacific Northwest; Rocky Mountain; Southern. Following \eqref{eq:omega_aggregate}, regional annual total carbon estimates for $\Omega_{\mathcal{J}}$ are given in Figure~\ref{fig:fia_regions_annual_totals}. This figure highlights the strong regional trends seen in Figures~\ref{fig:real_theta} and \ref{fig:real_delta}. Tabular estimates for trend $A_\mathcal{J}\theta_{\mathcal{J}}$ (Tg/year) and change $A_\mathcal{J}\Delta_{\mathcal{J}}$ (Tg) by FIA region are given in Table~\ref{tab:fia_regions_real_change} and again underscores patterns seen in Figure~\ref{fig:fia_regions_annual_totals}. Specifically, over the 14 year period, we see no significant change in carbon in the Northern region, significant carbon loss in the Pacific Northwest and Rocky Mountain regions, and significant carbon gain in the Southern region. 
\clearpage

\section{Discussion and summary}\label{sec:summary}

The simulation study and FIA data analysis demonstrate the proposed model's usefulness for delivering estimates for spatial and temporal extents with limited observations. The SAE model is assembled using components that are well documented in the statistical literature. In the current setting, the FH model provides a mechanism to learn from direct estimates, predictor variables, and spatially and temporally structured random effects. Casting this model into a Bayesian estimation framework provides additional opportunities to glean information from the data and survey design. For example, our proposed informative IG prior on $\sigma_{j,t}^2$ is an intuitive way to obtain potential information from the observed sample size $n_{j,t}$. The Bayesian framework also provides easy access to posterior distributions derived from the latent mean, e.g., total, trend, and change, for both individual areal units and aggregates of areal units. Further, inference from areal units with no observations is available via the latent mean's posterior predictive distribution, which allows us to make spatially and temporally complete estimates (e.g., for every county and year in the CONUS). This ability to use auxiliary information to adjust/inform direct estimates seems particularly useful for meeting evolving inferential demands placed on existing NFI programs.

As always, there is ``no free lunch.'' SAE models, including the one highlighted here, simply adjust or smooth the direct estimate. If there are issues with the direct estimate, then those issues persist in the model estimate. For example, the simulation study showed that those counties with small sample sizes had biased direct estimates and those biases might persist in the model estimates. However, if the adjacent areal units and/or times had more reliable data or larger sample sizes, then the structured random effects shrink the mean toward the average of the proximate neighbors, which could mitigate bias. In fact, one reason why CAR random effects are popular in SAE is because of this bias reduction in sparsely populated regions. The simulation study identified clear advantages to using the model, including improved accuracy and uncertainty quantification in most cases. 

Several properties and limitations of area-level FH models require further consideration. First, if all observations within an areal unit have the same value, there will be no direct estimate variance and, hence, that information is not available to train the model. Therefore, it is treated as missing although it provides valuable information. In our study, this occurred when all plots within a county and year had zero carbon and, in such cases, we relied on the TCC predictor and random effects to inform the posterior predictive distribution from which we draw inference. Second, the FH model is most often applied to continuous and Normally distributed variables, although there are transformations that might accommodate discrete count-like variables, see, e.g., \cite{Ghosh2022}. There are many forest variables that do not have Normal support or lend themselves well to transformation, and in such cases the FH model will not be suitable. Lastly, area-level models, as presented here, do not support prediction for units not included in the initial adjacency matrix. If inference for a new areal unit is needed, the unit must be added to the adjacency matrix and the model refit. If any of these limitations are prohibitive and NFI plot measurements are indexed in space and time, then one might consider unit-level models based upon point-referenced geostatistical models.   

NFIs provide critical information on the ecological health and economic viability of forests. NFI programs will continue to be asked for finer spatial and temporal resolution information on forest parameter status, trend, and change to support initiatives like forest carbon monitoring required by the UNFCCC. Financial circumstances limit most NFI programs' ability to collect sufficient data to meet these user needs, especially when estimates are provided by classical design-based methods. SAE models, like those proposed here, are needed to help bridge this gap and allow NFIs to provide high-quality and reliable estimates for key forest parameters.

\makeatletter
\renewcommand \thesection{S\@arabic\c@section}
\renewcommand\thetable{S\@arabic\c@table}
\renewcommand \thefigure{S\@arabic\c@figure}
\renewcommand \theequation{S\@arabic\c@equation}
\makeatother

\section*{Supporting information}\label{sec:supporting}

\section{Simulation study}\label{sec:simulation}
We follow a frequentist view for the simulation study, where a single population is generated using fixed and known values for parameters, then estimates for parameters are computed from each of a large number of independent samples, i.e., $R$ replicates, taken from the population. These estimates are then compared with population parameters using a set of measures that assess the estimators' bias, accuracy, and precision. Specifically, we are interested in assessing estimator quality for $\mu_{j,t}$ and derived parameters $\theta_j$ and $\Delta_j$. The four estimators considered are the direct (\ref{direct:mean}), Full model (\ref{mod:full}), Submodel 1 (\ref{mod:sub1}), and Submodel 2 (\ref{mod:sub2}).

\subsection{Simulated population and replicate samples}
To mimic qualities of the observed annual county-level data, we simulated a population comprising 7,809,952 point-referenced population units laid out in a 1-by-1 (km) regular grid over the CONUS land area. Each point-referenced population unit represents a possible sampling unit (i.e., FIA plot). At each population unit, we simulated a spatially and temporally correlated outcome value with dependence structure similar to the observed FIA plot carbon density (Mg/ha) data. Specifically, the outcome $y_t(\bell)$ at generic population unit location $\bell$ and time $t$ is given by 
\begin{align}
    y_t(\bell) &= \zeta_0 + u_t(\bell) + \zeta_1v_{TCC,t}(\bell) + \epsilon_t(\bell),\quad \epsilon_t(\bell)\iidsim N(0, \sigma_y^2),\nonumber \\
    u_t(\bell) &= u_{t-1}(\bell) + w_t(\bell), \quad w_t(\bell)\indsim GP(0, C(\cdot,\gamma)), \quad t = 1, 2, \ldots, T,
    \label{eq:sim_y}
\end{align}
where $\zeta_0$ and $\zeta_1$ are an intercept and slope coefficient, $u_t(\bell)$ is a space-time varying intercept, and $\epsilon_t(\bell)$ is a serially and spatially uncorrelated residual term following a mean zero Normal distribution with variance $\sigma^2_y$. The temporally independent $w_t(\bell)$ follows a mean zero GP with exponential covariance function $C(\cdot, \gamma) = \sigma^2_w \text{exp}(-\gamma||\bell - \bell'||)$, where $\sigma^2_w$ is the spatial variance, $||\bell - \bell'||$ is the Euclidean distance between two, possibly different, locations, and $\gamma$ is the decay parameter that controls the correlation. We set $u_0(\bell) = 0$.

To generate outcome patterns and relationships similar to the real FIA data, values for $v_{TCC,t}(\bell)$ were taken from the spatially and temporally coinciding TCC pixels and parameter values were set to $\zeta_0 = 1.5$, $\zeta_1 = 2$, $\sigma^2_y = 1,000$, $\sigma^2 = 10$, and $\gamma = 0.003$. These parameter values were chosen through exploratory regression analysis. The spatial decay parameter value produces a high correlation between adjacent counties (as seen in the real FIA data when aggregated to the county and year level). Also, to mimic observed forest and non-forest patterns, $y_t(\bell)$ was set to zero when $v_{TCC,t}(\bell) = 0$ and when the simulated $y_t(\bell)$ was less than zero. 

\begin{figure}[H]
\centering
\includegraphics[trim={0 1cm 0 1cm},clip,width=\textwidth]{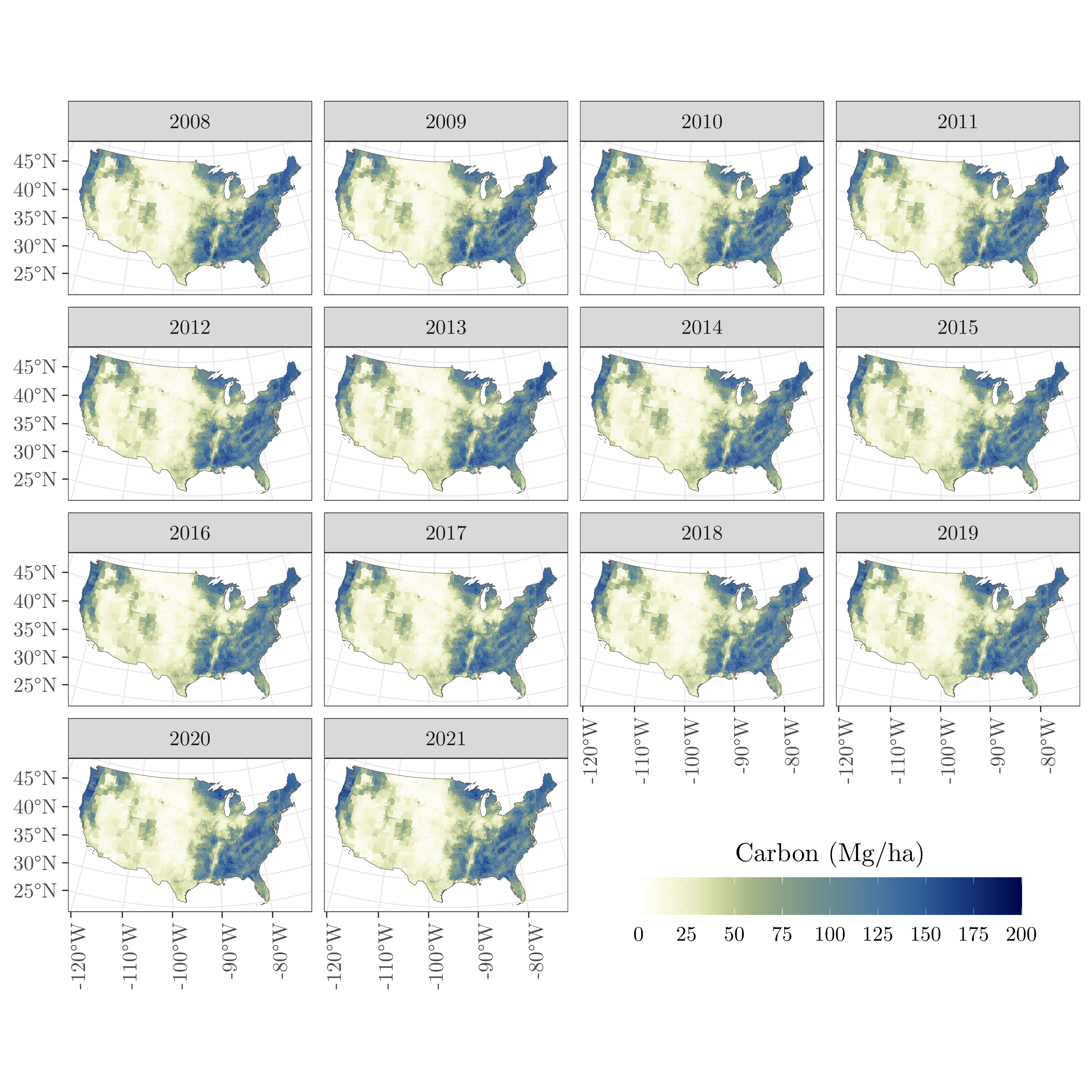}
\caption{Simulated population forest carbon density $\mu_{\text{true}, t, j}$ (Mg/ha).}
\label{fig:sim_mu}
\end{figure}

\begin{figure}[ht!]
\centering
\includegraphics[trim={0 4.75cm 0 4.5cm},clip,width=\textwidth]{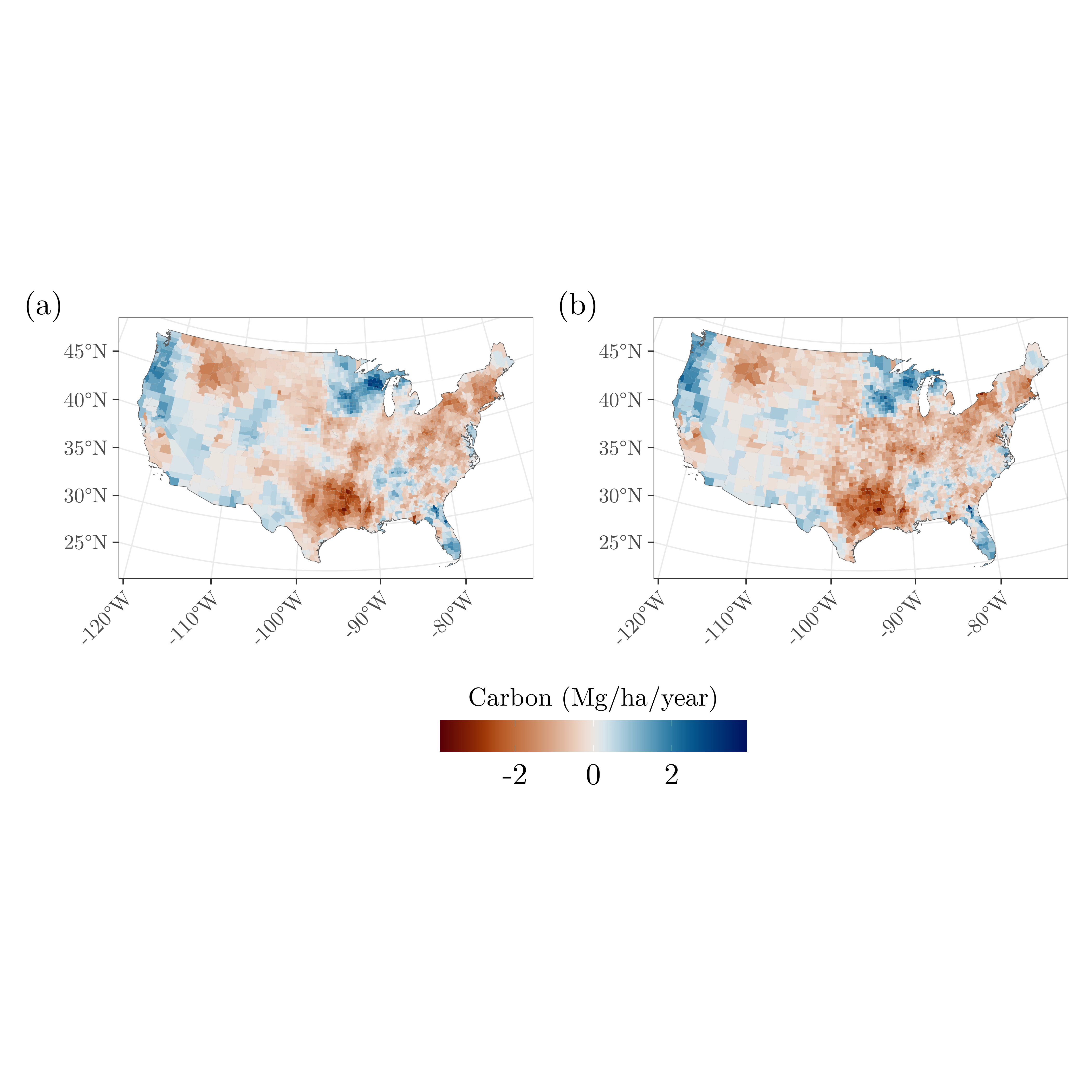}
\caption{(a) Simulated population linear slope trend in forest carbon density $\theta_j$ (Mg/ha/year). (b) Full model estimates for $\theta_j$ using the first replicate's sample data.}
\label{fig:sim_mu_theta}
\end{figure}

\begin{figure}[ht!]
\centering
\includegraphics[trim={0 4.75cm 0 4.5cm},clip,width=\textwidth]{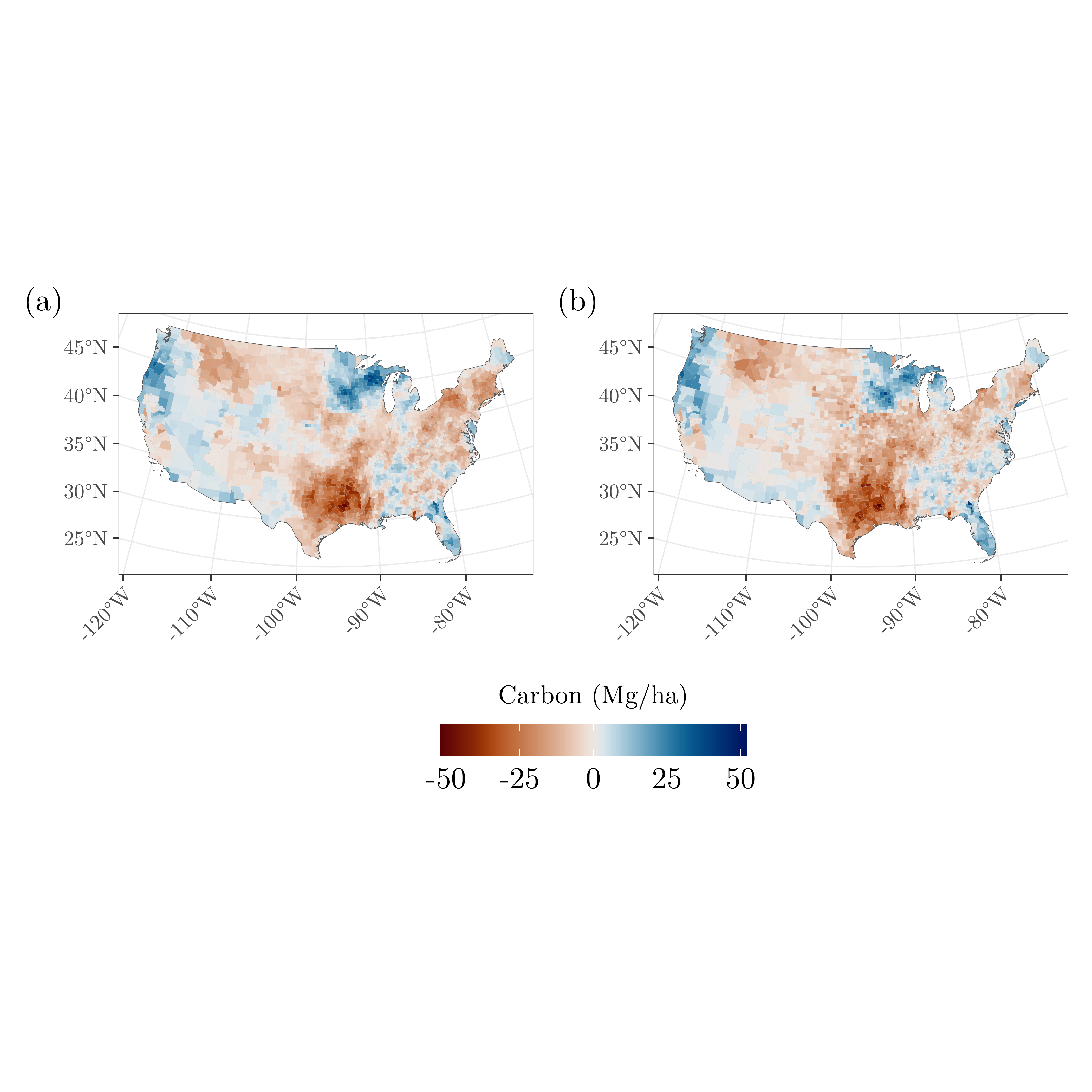}
\caption{(a) Simulated population change in forest carbon density $\Delta_j$ (Mg/ha) between $t_1$=2008 and $t_2$=2021. (b) Full model estimates for $\Delta_j$ using the first replicate's sample data.}
\label{fig:sim_mu_delta}
\end{figure}

Simulated population values from (\ref{eq:sim_y}) were averaged within each county and year and serve as the ``true'' mean $\mu_{true,j,t}$ shown in Figure~\ref{fig:sim_mu}. This true mean was used to calculate $\theta_{true,j}$ using (\ref{eq:mu_theta}) between 2008 and 2021 and $\Delta_{true,j}$ using (\ref{eq:mu_delta}) with $t_1$ = 2008 and $t_2$ = 2021. The resulting true county-level trend and change are shown in Figures~\ref{fig:sim_mu_theta}(a) and \ref{fig:sim_mu_delta}(a), respectively. 

Following (\ref{direct:mean}) and (\ref{direct:var}), annual county-level direct estimates were generated from each of $R$=100 sample replicates. Each replicate comprises a simple random sample from the population, with the sample size for each county and year set to that of the FIA data (Figure~\ref{fig:sample_size}). Although the proposed models can accommodate sample sizes of zero (as demonstrated in the FIA data analysis Section~\ref{sec:real}), comparisons consider only those counties with two or more plots so that the direct estimator's uncertainty quantification can be compared with the models'. 

\subsection{Comparison among estimators}
Given $\mu_{true,j,t}$ and $R$ estimates from each estimator, we consider measures of bias, root mean squared error (RMSE), empirical confidence and credible interval coverage rates, and width between the lower and upper 95\% confidence and credible interval bounds, as defined below.

Systematic difference between estimate and truth is evaluated using
\begin{equation}
    \text{Bias}_{j,t,l} = \frac{\sum^{R}_{r = 1} \left(\mu_{j,t,l,r} - \mu_{\text{true},j,t}\right)}{R},
\label{eq:bias}
\end{equation}
where $\mu_{j,t,l,r}$ is the $l^{\text{th}}$ estimator's point estimate given sample data from the $r^\text{th}$ replicate. The point estimate for the candidate models is the posterior mean of $\mu_{j,t}$ and $\hat{\mu}_{j,t}$ for the design-based estimator.

Average accuracy is evaluated using   
\begin{equation}
    \text{RMSE}_{j,t,l} = \sqrt{\frac{\sum^{R}_{r = 1} \left(\mu_{j,t,l,r} - \mu_{\text{true},j,t}\right)^2}{R}}.
\label{eq:rmse}
\end{equation}

Average empirical coverage rate is evaluated using 
\begin{equation}
    \text{Cover}_{j,t,l} = \frac{\sum^{R}_{r = 1} I\left(\mu^L_{j,t,l,r} \le \mu_{\text{true},j,t} \le \mu^U_{j,t,l,r}\right)}{R},
    \label{eq:cover}
\end{equation}
where $I(\cdot)$ is an indicator function, and $\mu^L_{j,t,l,r}$ and $\mu^U_{j,t,l,r}$ are the lower and upper uncertainty quantification bounds. Here we consider the 95\% credible interval bounds for the candidate models and 95\% confidence intervals for the design-based estimator.

Average precision is evaluated using 
\begin{equation}
    \text{Width}_{j,t,l} = \frac{\sum^{R}_{r = 1} \left(\mu^U_{j,t,l,r} - \mu^L_{j,t,l,r}\right)}{R}.
    \label{eq:width}
\end{equation}

\begin{figure}[ht!]
\centering
\includegraphics[trim={0 0cm 0 0cm},clip,width=\textwidth]{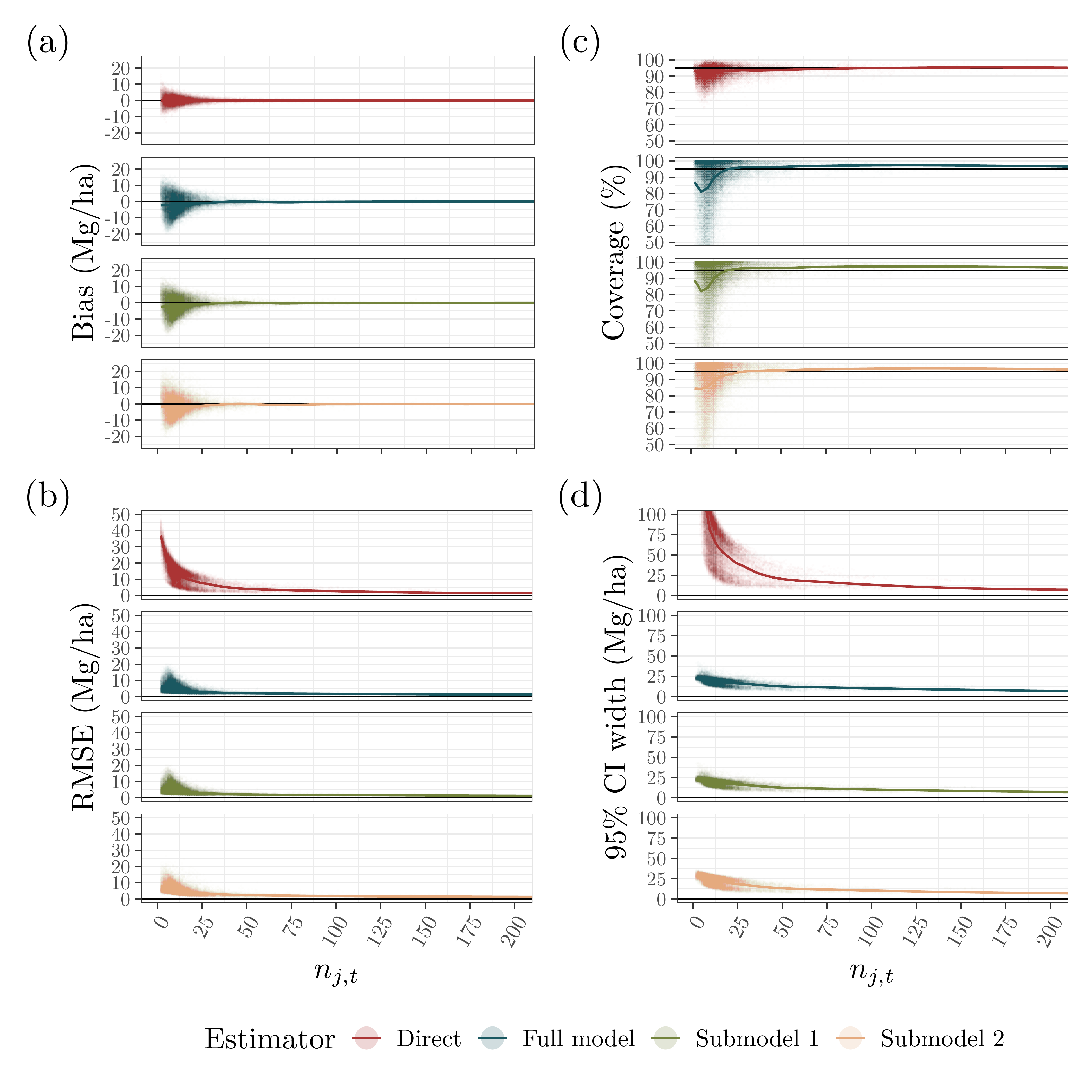}
\caption{Performance of estimators for $\mu_{j,t}$ from the simulation analysis. Each point value represents the average over $R$ replicates for a county and year. A loess line is added to indicate point scatter trend across sample size ($n_{j,t}$).}\label{fig:sim_results_mu}
\end{figure}

These performance measures are summarized in Figure~\ref{fig:sim_results_mu}, where each point is a county and year combination (i.e., there are $N$ points in each figure sub-panel). County and year combinations are ordered along the $x$-axis by increasing sample size. 

Figure~\ref{fig:sim_results_mu}(a) shows all estimators can be biased when sample size is small, e.g., $n_{j,t} \lesssim 50$, and this bias can increase as sample size decreases. For some of the smallest sample sizes, e.g., $n_{j,t}=2$, the direct estimate for a given county and year can produce bias as large as approximately $\pm10$ (Mg/ha). The models can yield slightly greater bias for small sample sizes, with Submodel 2 exhibiting the largest bias among the estimators. 

Figure~\ref{fig:sim_results_mu}(b) shows the direct estimator has larger RMSE (poorer accuracy) relative to the models when $n_{j,t} \lesssim 100$ and particularly so when $n_{j,t} \lesssim 25$. Figure~\ref{fig:sim_results_mu}(c) shows the direct estimator achieves close to the expected 95\% confidence interval coverage rate even for small samples. In contrast, the models demonstrate dramatically lower than expected credible interval coverage rates when the sample size is small. As shown in Figure~\ref{fig:sim_results_mu}(d), the direct estimator's confidence interval coverage rate of $\sim$95\% reflects its extremely large confidence interval widths. Conversely, the models' relatively poor credible interval coverage rates are due to overly narrow credible intervals.

\begin{figure}[ht!]
\centering
\includegraphics[trim={0 0cm 0 0cm},clip,width=\textwidth]{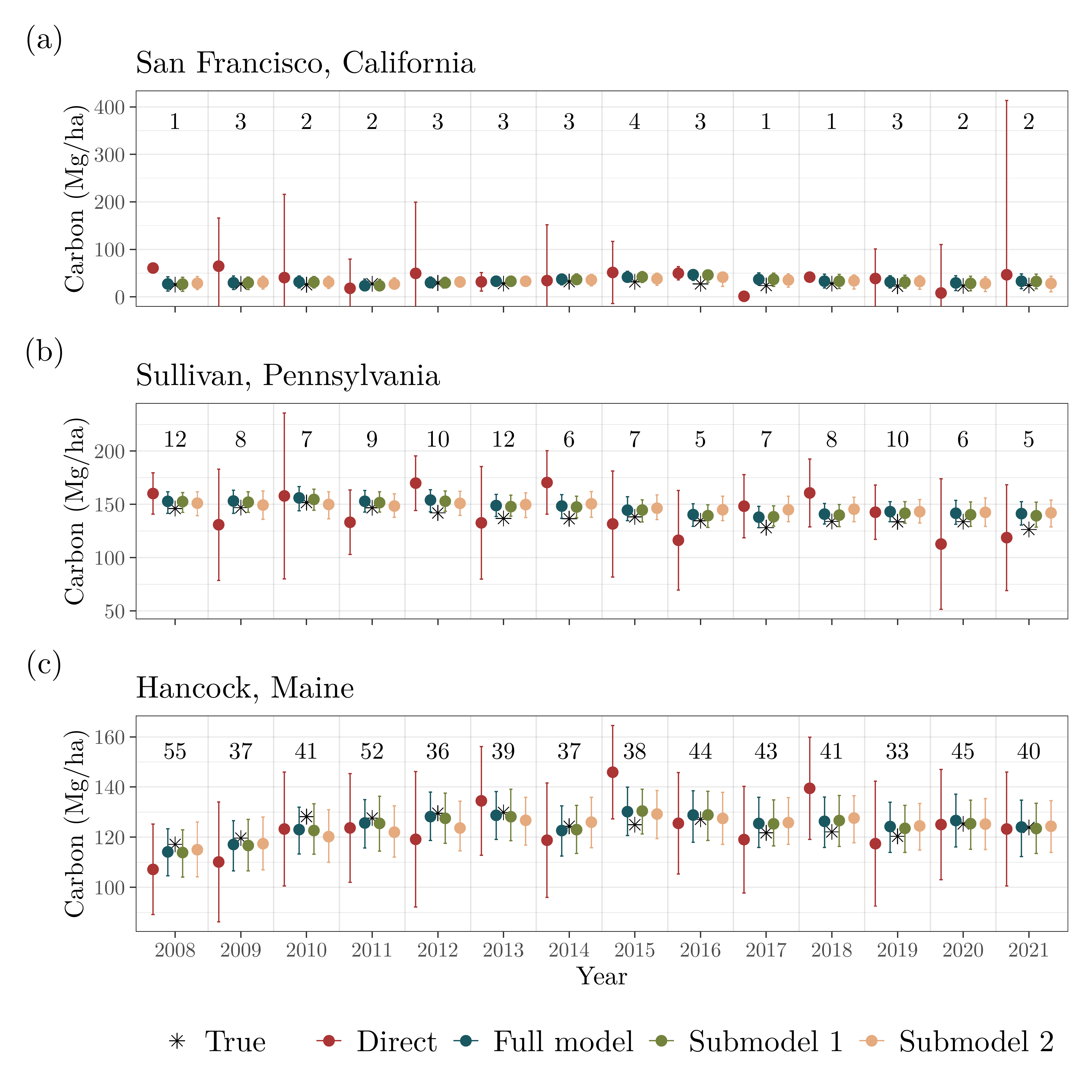}
\caption{Simulated $\mu_{true,j,t}$ values and estimates based on sample data from the first replicate. Sample size $n_{j,t}$ for each county and year is given across the top of each subpanel. Estimate means and medians are shown as points with 95\% confidence and credible interval bars for direct and model estimates, respectively. When the sample size is one, confidence intervals are not available for the direct estimate.}\label{fig:sim_results_mu_ex}
\end{figure}

Figure~\ref{fig:sim_results_mu_ex} shows estimates for three counties based on the first replicate's sample data, provides some insight into the summaries given in Figure~\ref{fig:sim_results_mu}, and demonstrates how estimates compare for low, moderate, and moderately large sample sizes (figures for all $J$ counties are provided in the Supplemental Material). Estimates for San Francisco, CA, are given in Figure~\ref{fig:sim_results_mu_ex}(a). For small, urban counties like San Francisco that have low forest carbon density values, little information is provided by direct estimates due to small sample sizes, and model estimates are informed by the TCC predictor variable and direct estimates in adjacent years and counties. In these cases, our results routinely show the models provide accurate estimates, but their credible intervals are often too narrow (i.e., do not cover the true value at the expected rate). This trend is captured in Figures~\ref{fig:sim_results_mu}(b-d). In contrast, model estimates for counties with moderately large sample sizes, such as Sullivan, PA, Figure~\ref{fig:sim_results_mu_ex}(b), and larger samples sizes, such as Hancock, ME, Figure~\ref{fig:sim_results_mu_ex}(c), have improved accuracy and precision over direct estimates.

\begin{figure}[ht!]
\centering
\includegraphics[trim={0 0cm 0 0cm},clip,width=\textwidth]{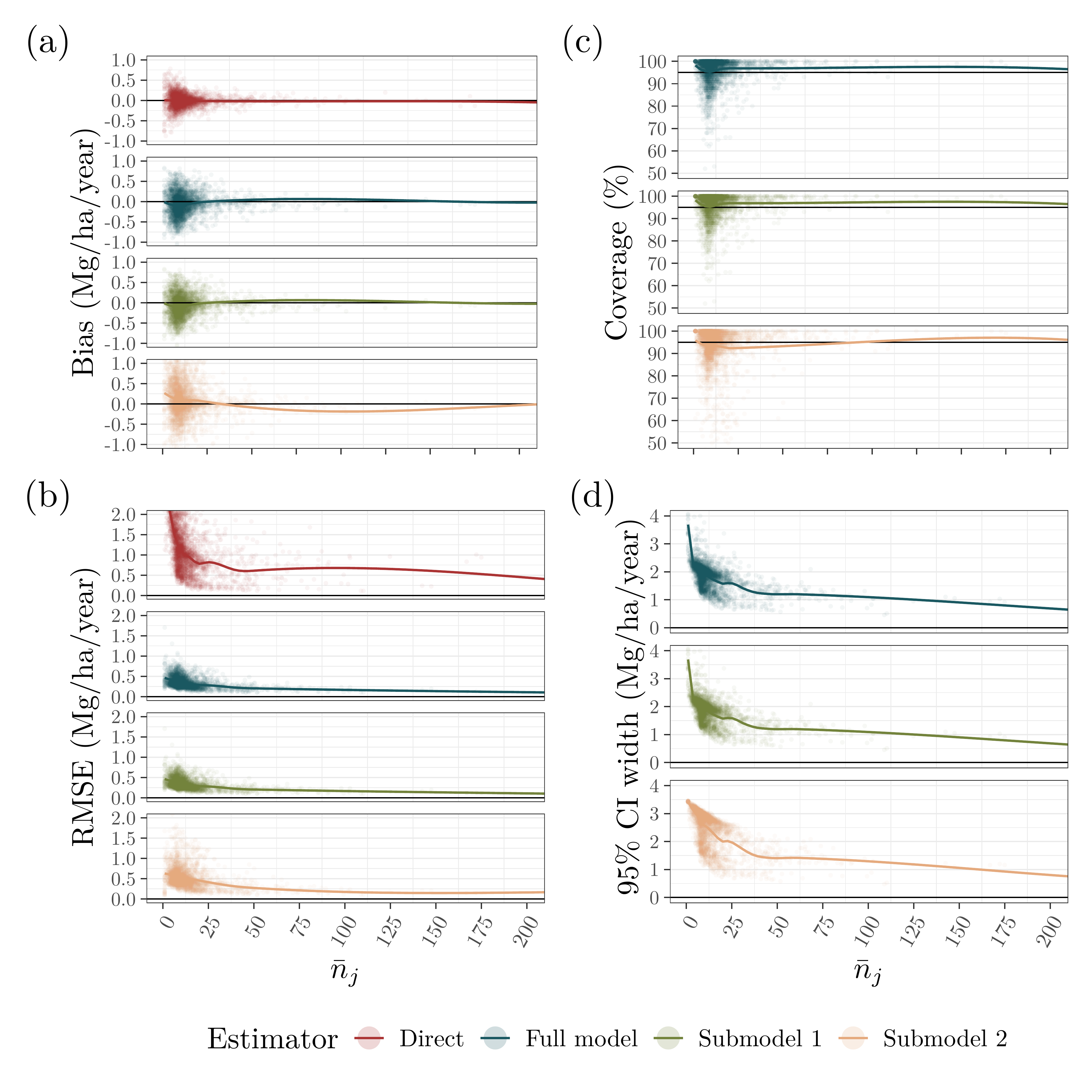}
\caption{Simulation analysis estimator performance for $\theta_j$. Point values are the average over $R$ replicates for a county. A loess line is added to indicate point scatter trend across average sample size values ($\bar{n}_{j} = \frac{1}{T} \sum_{t = 1}^T n_{j,t}$).}\label{fig:sim_results_mu_theta}
\end{figure}

\begin{figure}[ht!]
\centering
\includegraphics[trim={0 0cm 0 0cm},clip,width=\textwidth]{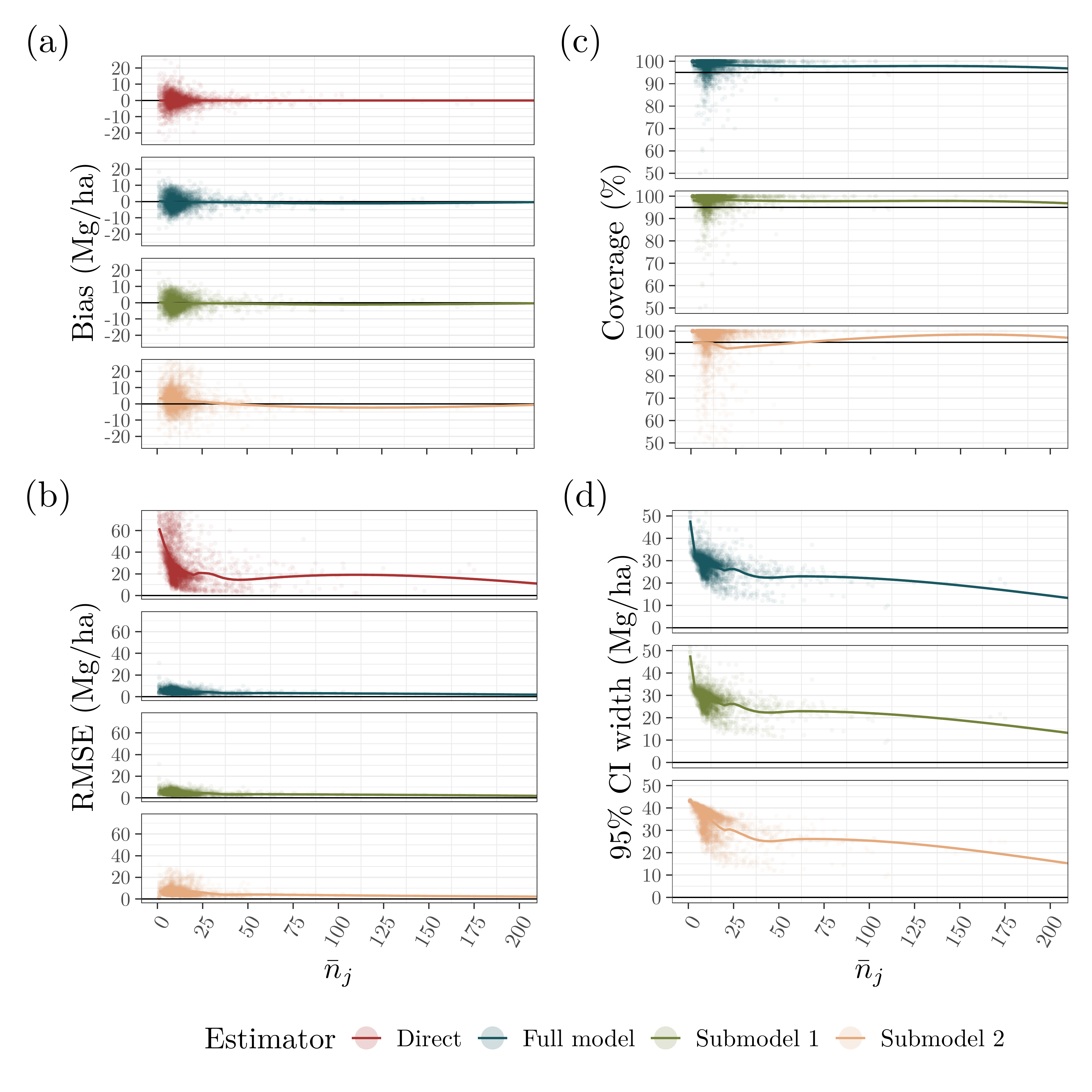}
\caption{Simulation analysis estimator performance for $\Delta_j$. Each point value represents the average over $R$ replicates for a county. A loess line is added to indicate point scatter trend across average sample size values ($\bar{n}_{j} = \frac{1}{T}\sum_{t = 1}^T n_{j,t}$).}\label{fig:sim_results_mu_delta}
\end{figure}

The simulated population values for the trend $\theta_j$ (\ref{eq:mu_theta}) and change $\Delta_j$ (\ref{eq:mu_delta}) parameters are shown in Figures~\ref{fig:sim_mu_theta} and \ref{fig:sim_mu_delta}, respectively. For these parameters, the models are able to deliver improved accuracy and precision with bias comparable to the direct estimator and coverage rate closer to the expected 95\%, see Figures~\ref{fig:sim_results_mu_theta} and \ref{fig:sim_results_mu_delta}, respectively. To visualize this accuracy spatially, the parameters' true values shown in Figures~\ref{fig:sim_mu_theta}(a) and \ref{fig:sim_mu_delta}(a) can be compared with estimates based on the first replicate's sample data given in Figures~\ref{fig:sim_mu_theta}(b) and \ref{fig:sim_mu_delta}(b).

Returning to Figure~\ref{fig:sim_results_mu}, we observe clear differences between the direct and model estimators; however, it is difficult to discern performance differences between the models. Following Section~\ref{sec:waic}, WAIC was estimated for each candidate model and each replicate. When comparing WAIC scores, Submodel 1 and the Full model estimates are comparable, and both are substantially better than Submodel 2 (Table~\ref{tab:sim_waic}). As specified in (\ref{mod:sub1}), Submodel 1 has a space- and time-varying intercept but a spatially and temporally invariant regression coefficient; hence, Submodel 1 is consistent with the data simulation model (\ref{eq:sim_y}). We therefore might expect Submodel 1 to outperform the Full model. However, given the flexibility of the Full model, it is not surprising that the information criterion does not show a substantial difference between Submodel 1 and the Full model. Similar results presented in \cite{Doser2024} showed that WAIC could not differentiate between a space-varying intercept and a SVC model when simulated data came from a space-varying intercept model.

\begin{table}[ht!]
\begin{center}
\begin{tabular}{lccc}
\toprule
&\multicolumn{3}{c}{Candidate models} \\
\cmidrule(lr){2-4} 
Parameter & Submodel 2 & Submodel 1 & Full model\\
\midrule
$\widehat{\text{elpd}}_{\text{WAIC}}$ &-164857.2 (157.0)&-164177.9 (156.2)&-164164.1 (155.3)\\
$\widehat{p}_{\text{WAIC}}$ &9630.5  (96.8)&8477.8 (104.9)&8454.4 (103.7)\\
WAIC &329714.3 (313.9)&328355.7 (312.3)&328328.1 (310.7)\\
\bottomrule
\end{tabular}
\caption{Simulated data estimates for WAIC and associated statistics. Values are the average over the replicates with the standard deviation given in parentheses.}\label{tab:sim_waic}
\end{center}
\end{table} 

\begin{table}[ht!]
\begin{center}
\begin{tabular}{lccc}
\toprule
&\multicolumn{3}{c}{Candidate models} \\
\cmidrule(lr){2-4} 
Parameter & Submodel 2 & Submodel 1 & Full model\\
\midrule
$\beta_0$ &69.34 (69.03, 69.73)&68.43 (67.79, 70.55)&66.65 (65.72, 67.27)\\
$\beta_{TCC}$ &46.66 (46.42, 46.92)&48.33 (47.94, 48.82)&47.65 (46.17, 48.90)\\
$\alpha_{\eta^{t}_0}$ &0.9192 (0.9119, 0.9265)&-&-\\
$\sigma^2_{\eta^{t}_0}$ &82.79 (77.88, 87.10)&-&-\\
$\rho_{\eta^{st}_0}$  &-&0.9984 (0.9979, 0.9989)&0.9989 (0.9982, 0.9993)\\
$\alpha_{\eta^{st}_0}$  &-&0.8918 (0.8811, 0.9053)&0.8773 (0.8645, 0.8893)\\
$\sigma^2_{\eta^{st}_0}$ &-&111.74 (103.21, 121.31)&99.13 90.36 109.08)\\
$\rho_{\eta^s_{TCC}}$  &-&-&0.9998 (0.9986, 1.00)\\
$\sigma^2_{\eta^{s}_{TCC}}$ &-&-&9.75 (7.47, 13.17)\\
$\sigma^2_{\epsilon}$  &7.06 (5.82, 8.51)&9.66 (8.56, 10.94)&10.07 (8.83, 11.37)\\
\bottomrule
\end{tabular}
\caption{Simulated data analysis parameter estimates for candidate models using first replicate's sample data. Estimates are posterior medians with 95\% credible interval given in parentheses.}\label{tab:sim_ests}
\end{center}
\end{table} 

Table~\ref{tab:sim_ests} provides parameter estimates for each candidate model based on the first replicate's sample data, which are representative of those generated from the other $R-1$ sample datasets. Although not of direct interest to model comparison, these estimates give a sense of the process parameter values and, in particular, the strength of the spatial and temporal correlation parameters. The strong spatial and temporal correlations seen in the Submodel 1 and Full model intercept random effect are also observed in the FIA data analysis presented in Section~\ref{sec:real}. Given the similarities between the simulated and FIA data, we gain some assurance that the estimators' qualities explored in Figures~\ref{fig:sim_results_mu}, \ref{fig:sim_results_mu_theta}, and \ref{fig:sim_results_mu_delta} are transferable to the FIA data analysis.

\begin{figure}[ht!]
\centering
\includegraphics[trim={0 0cm 0 0cm},clip,width=\textwidth]{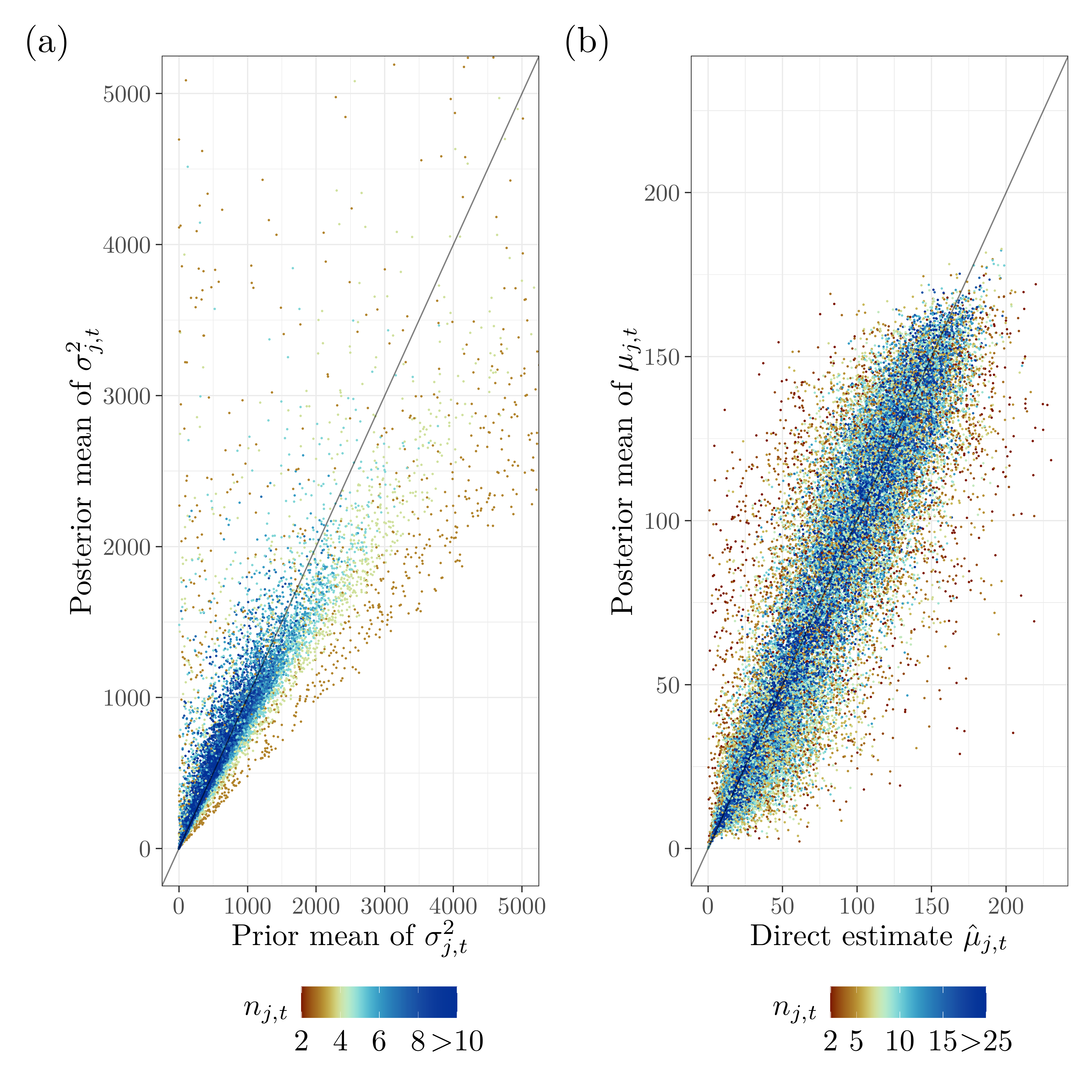}
\caption{Full model parameter estimates using first replicate's sample data. (a) $\sigma^2_{j, t}$ prior mean versus its posterior mean. (b) $\mu_{j,t}$ direct estimate versus its posterior mean. Points colored by sample size. }\label{fig:sim_sigmaSq_prior}
\end{figure}

By design, the hierarchical prior for $\sigma^2_{j,t}$ given in (\ref{sec:model}) provides a varied effect on its posterior distribution. This effect is directly related to the number of observations ($n_{j,t}$) used to calculate the associated direct estimate, whereby more weight is placed on the prior mean as $n_{j,t}$ increases. As shown in Figure~\ref{fig:sim_sigmaSq_prior}(a), for $n_{j,t} \gtrsim 8$, the posterior mean more closely reflects the prior mean, (blue points along the 1:1 line) while discrepancy between prior and posterior means is clear for $n_{j,t} \lesssim 4$ (yellow to red points away from the 1:1 line). This discrepancy stems from learning that occurs from other information present in the data that the model essentially weights more heavily, including neighboring direct estimates, predictor variables, and structured random effects. This balance between information sources conditional on $n_{j,t}$ is also seen in Figure~\ref{fig:sim_sigmaSq_prior}(b), where we compare $\mu_{j,t}$'s direct estimate to its estimated posterior distribution mean. Here, again, as $n_{j,t}$ increases, more information is drawn from the direct estimate (blue points close the 1:1 line), and when $n_{j,t}$ is small, more information is drawn from other sources and estimates can be quite different from the direct estimate (red points away from the 1:1 line).  

\section{Supporting figures and tables}\label{sec:supporting_figures_tables}

\begin{figure}[ht!]
\centering
\includegraphics[trim={0 3.25cm 0 1cm},clip,width=\textwidth]{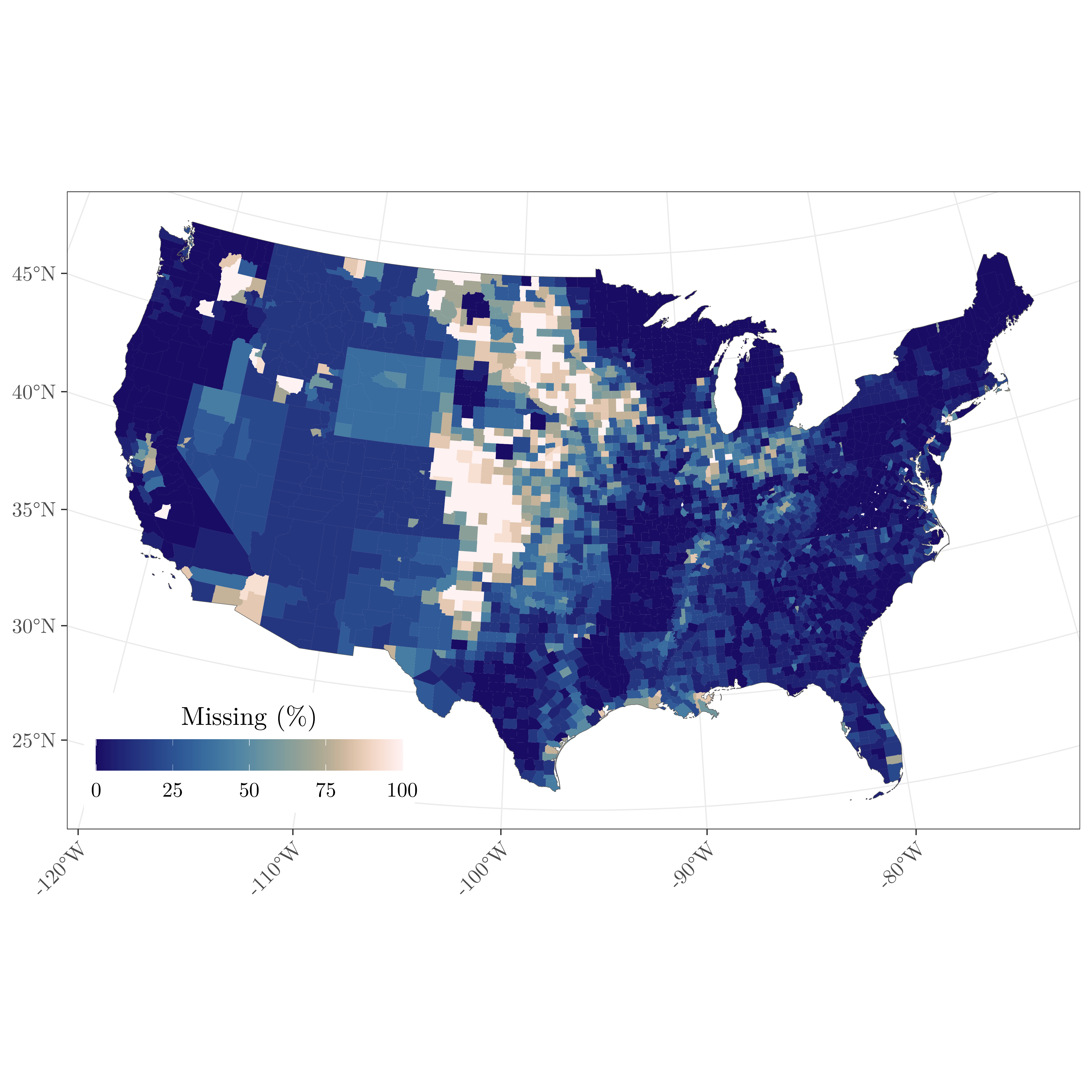}
\caption{Percent of missing direct estimates over the 14 year study period.}
\label{fig:missing}
\end{figure}

\begin{figure}[ht!]
\centering
\includegraphics[trim={0 1cm 0 0cm},clip,width=\textwidth]{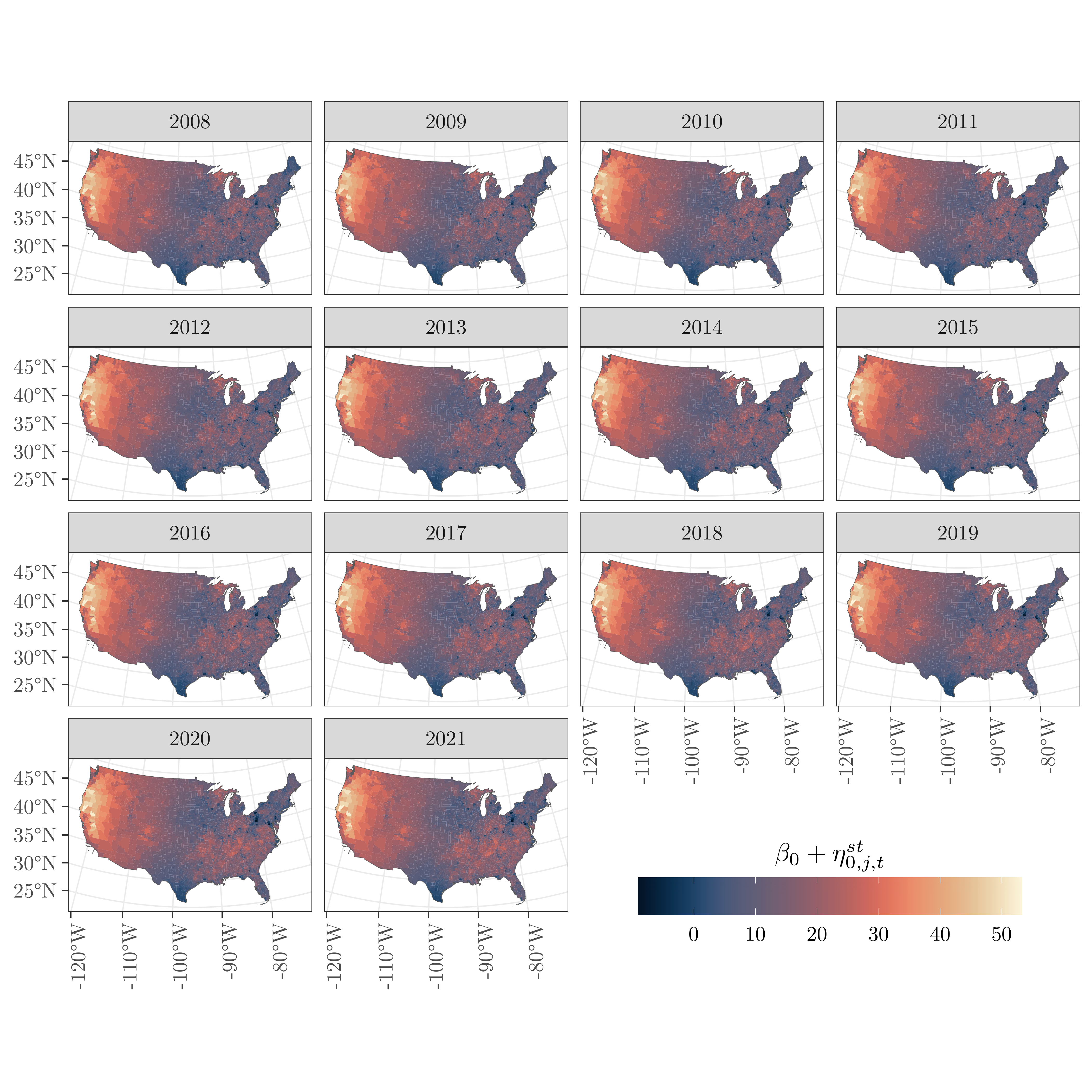}
\caption{Posterior median of the Full model's time- and space-varying intercept $\beta_{0}+\eta^{st}_{0,j,t}$ fit with FIA data.}\label{fig:full_svi}
\end{figure}

\begin{figure}[ht!]
\centering
\includegraphics[trim={0 1cm 0 1cm},clip,width=\textwidth]{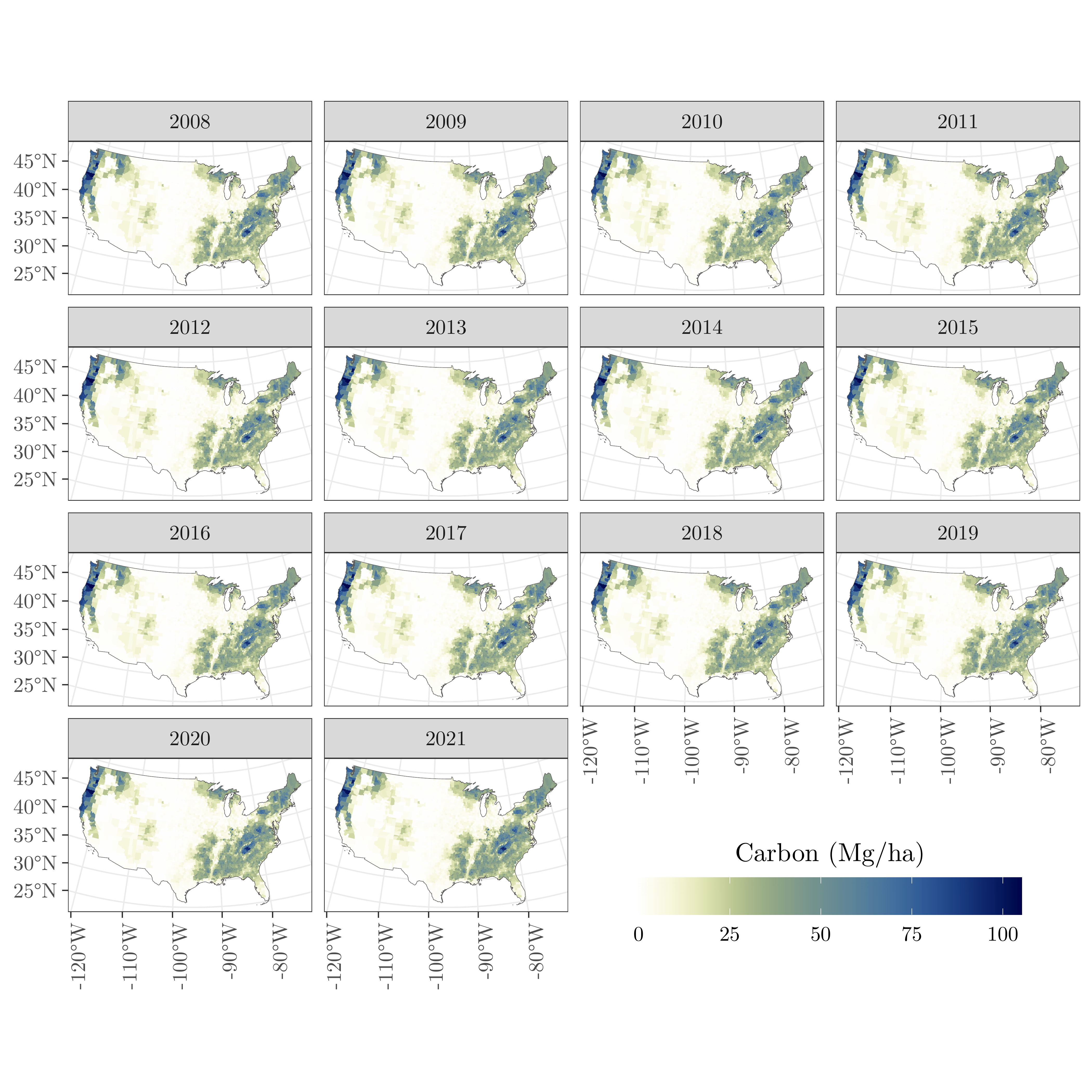}
\caption{Full model estimates for $\mu_{j,t}$'s posterior mean fit with FIA data.}\label{fig:real_carbon_mean}
\end{figure}

\begin{figure}[ht!]
\centering
\includegraphics[trim={0 1cm 0 1cm},clip,width=\textwidth]{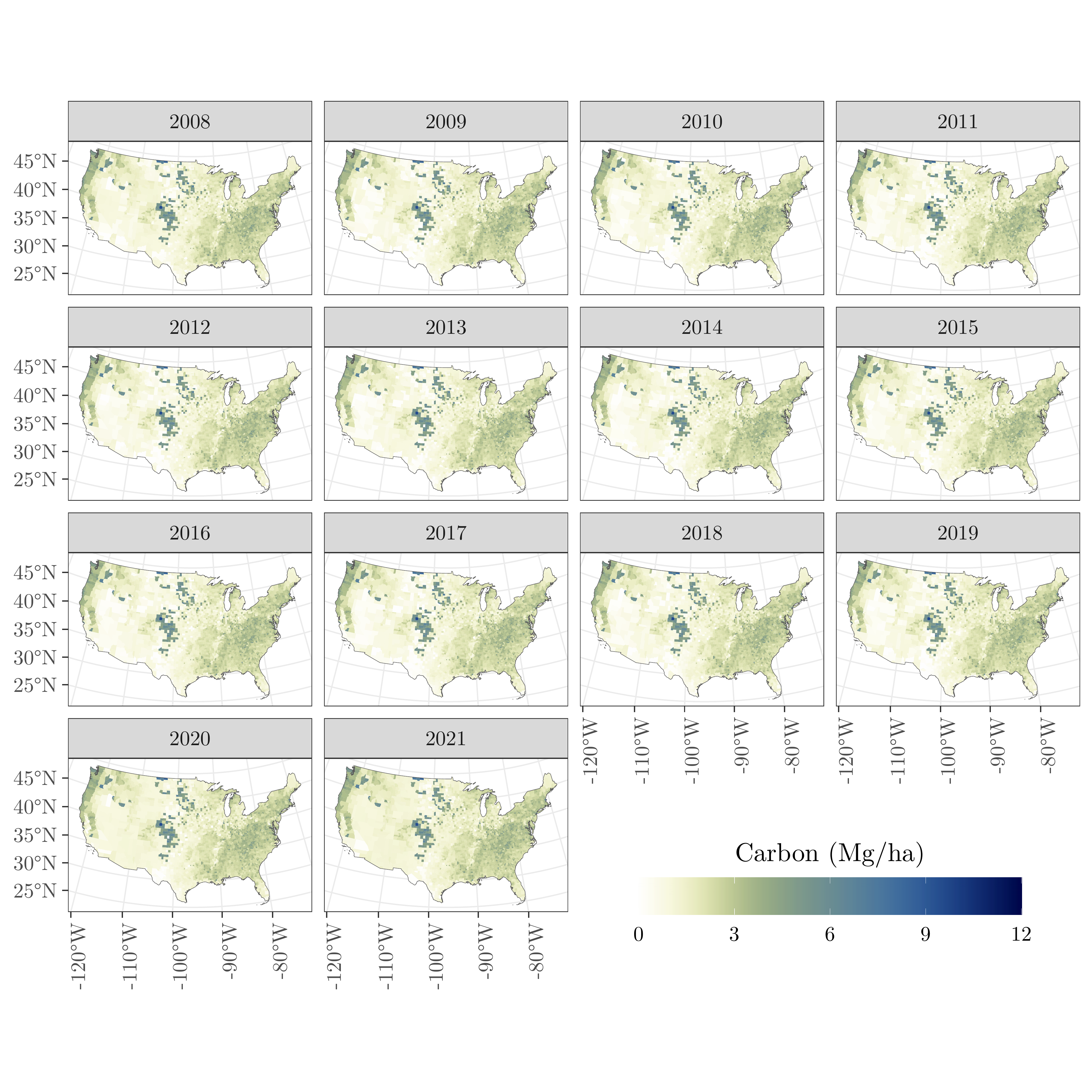}
\caption{Full model estimates for $\mu_{j,t}$'s posterior standard deviation fit with FIA data.}\label{fig:real_carbon_sd}
\end{figure}

\begin{table}[ht!]
\begin{center}
\begin{tabular}{llcllc}
\toprule
\multicolumn{3}{c}{Decreasing } & \multicolumn{3}{c}{Increasing} \\
\cmidrule(lr){1-3} 
\cmidrule(lr){4-6} 
State & County & Carbon &State & County & Carbon\\
\cmidrule(lr){1-3} 
\cmidrule(lr){4-6} 
California&Lake&-1.25 (-1.62, -0.88)&Alabama&Escambia&0.92 (0.67, 1.20)\\
Florida&Calhoun&-1.21 (-1.63, -0.76)&Georgia&Cobb&0.87 (0.58, 1.17)\\
California&Mariposa&-1.19 (-1.57, -0.77)&Tennessee&Rhea&0.86 (0.59, 1.15)\\
Florida&Gulf&-0.94 (-1.40, -0.52)&Mississippi&Hancock&0.85 (0.55, 1.14)\\
California&Tuolumne&-0.76 (-1.04, -0.48)&Alabama&Dallas&0.83 (0.58, 1.07)\\
California&Shasta&-0.71 (-0.98, -0.44)&Mississippi&Harrison&0.82 (0.52, 1.10)\\
Pennsylvania&Allegheny&-0.67 (-0.92, -0.42)&Alabama&Conecuh&0.82 (0.54, 1.10)\\
California&Napa&-0.66 (-1.00, -0.34)&Alabama&Greene&0.81 (0.54, 1.07)\\
Virginia&Falls Church&-0.64 (-1.11, -0.21)&Mississippi&Stone&0.81 (0.55, 1.08)\\
Ohio&Carroll&-0.63 (-0.93, -0.34)&Georgia&Polk&0.80 (0.55, 1.07)\\
\bottomrule
\end{tabular}
\caption{Ten largest decreasing and increasing estimates for carbon density trend $\theta_j$ (Mg/ha/year) from 2008 to 2021 across the CONUS. Estimates are posterior medians with 95\% credible interval values given in parentheses.}\label{tab:real_largest_theta}
\end{center}
\end{table} 

\begin{table}[ht!]
\begin{center}
\begin{tabular}{llcllc}
\toprule
\multicolumn{3}{c}{Decreasing } & \multicolumn{3}{c}{Increasing} \\
\cmidrule(lr){1-3} 
\cmidrule(lr){4-6} 
State & County & Carbon &State & County & Carbon\\
\cmidrule(lr){1-3} 
\cmidrule(lr){4-6} 
Idaho&Idaho&-0.92 (-1.41, -0.45)&Maine&Aroostook&0.71 (0.35, 1.05)\\
California&Siskiyou&-0.92 (-1.33, -0.48)&Minnesota&St. Louis&0.64 (0.34, 0.96)\\
California&Shasta&-0.7 (-0.98, -0.44)&Maine&Piscataquis&0.57 (0.31, 0.83)\\
Washington&Okanogan&-0.69 (-1.00, -0.36)&Maine&Penobscot&0.49 (0.29, 0.68)\\
Oregon&Douglas&-0.47 (-0.87, -0.11)&Maine&Washington&0.46 (0.26, 0.64)\\
California&Mariposa&-0.45 (-0.60, -0.29)&Maine&Somerset&0.44 (0.22, 0.66)\\
California&Tuolumne&-0.45 (-0.61, -0.28)&Alabama&Baldwin&0.29 (0.19, 0.40)\\
California&Lake&-0.43 (-0.56, -0.30)&Maine&Hancock&0.27 (0.14, 0.40)\\
California&Trinity&-0.43 (-0.67, -0.20)&Minnesota&Cook&0.25 (0.08, 0.43)\\
California&Fresno&-0.42 (-0.73, -0.11)&Minnesota&Itasca&0.24 (0.07, 0.42)\\
\bottomrule
\end{tabular}
\caption{Ten largest decreasing and increasing estimates of total carbon trends $A_j\theta_j$ (Tg/year) from 2008 to 2021 across the CONUS, where $A_j$ is the area of county $j$ in hectares. Estimates are posterior medians with 95\% credible interval values given in parentheses.}\label{tab:real_largest_theta_carbon}
\end{center}
\end{table} 

\begin{figure}[ht!]
\centering
\includegraphics[trim={0 1cm 0 1cm},clip,width=\textwidth]{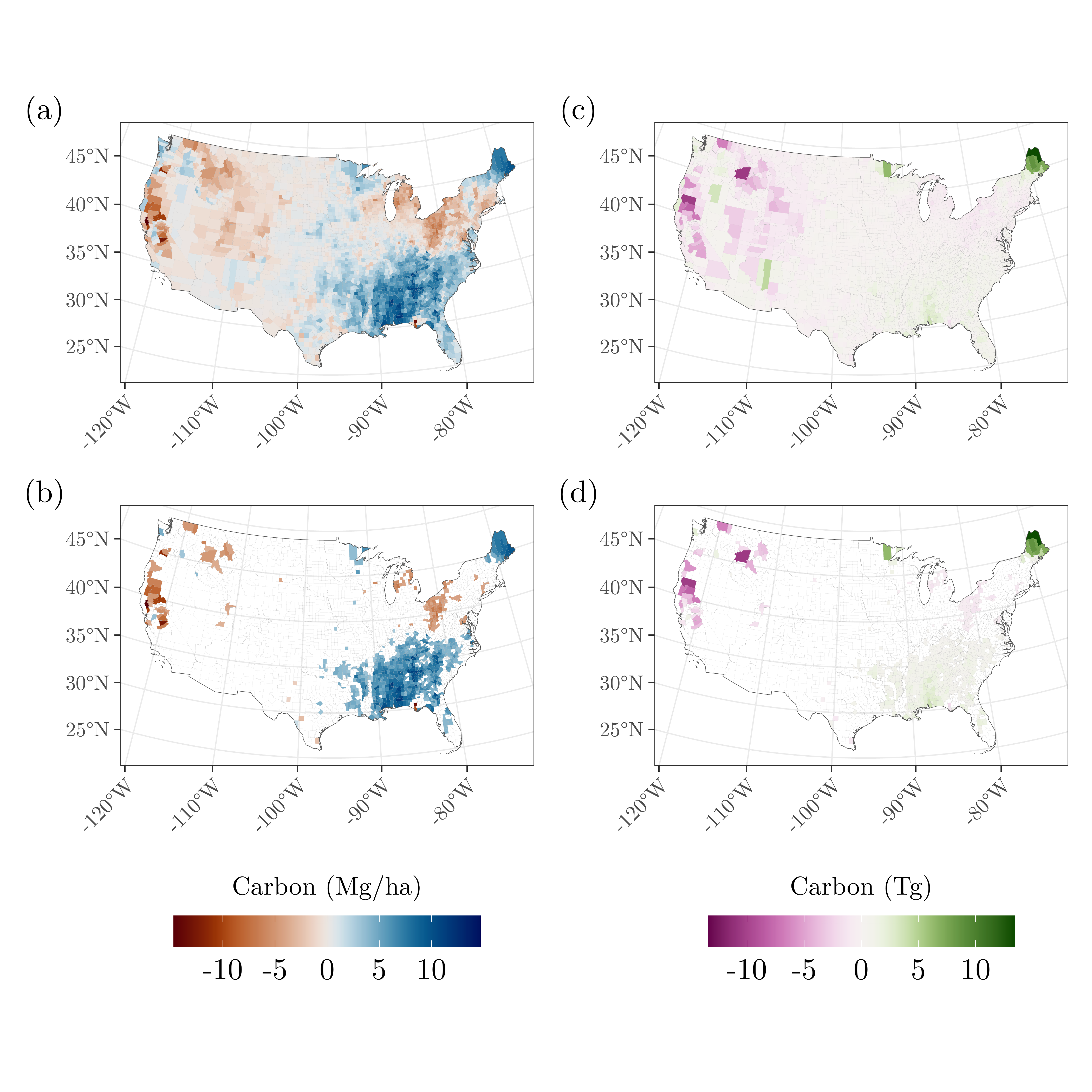}
\caption{(a) Estimated change in forest carbon density between 2021 and 2008 $\Delta_j$ (Mg/ha). Values are each county's posterior distribution median. (b) Counties from (a) that have posterior distribution 95\% credible intervals that exclude zero. (c) Estimated change in forest carbon total $A_j\Delta_j$ (Tg), where $A_j$ is the $j^{\text{th}}$ county's area in hectares. (d) Counties from (c) that have posterior distribution 95\% credible intervals that exclude zero. }\label{fig:real_delta}
\end{figure}

\begin{figure}[ht!]
\centering
\includegraphics[trim={0 0cm 0 0cm},clip,width=\textwidth]{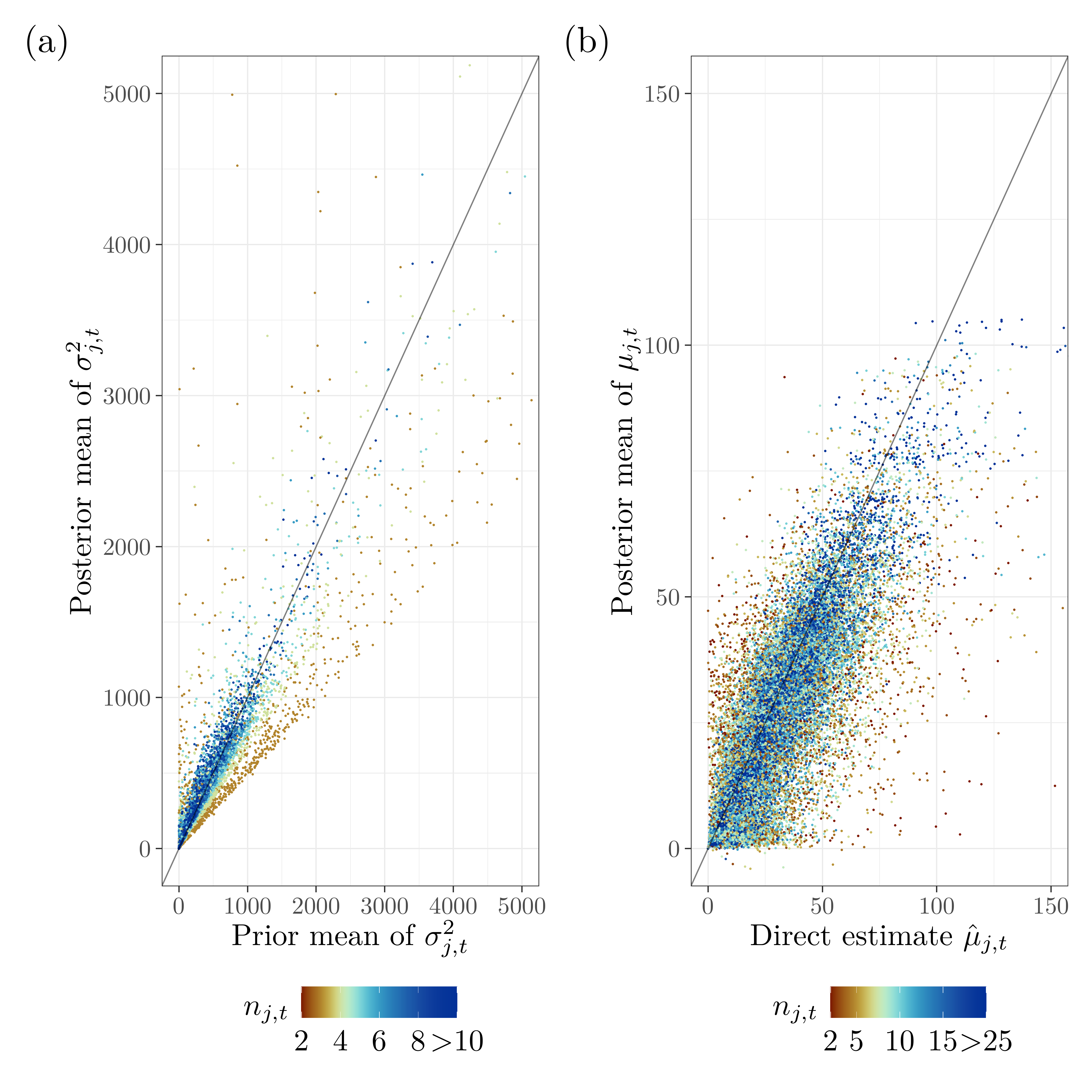}
\caption{FIA data Full model estimates. (a) $\sigma^2_{j, t}$ prior mean versus its posterior mean. (b) $\mu_{j,t}$ direct estimate versus its posterior mean. Points colored by sample size.}\label{fig:real_sigmaSq_prior}
\end{figure}

\begin{figure}[ht!]
\centering
\includegraphics[trim={0 1cm 0 1cm},clip,width=\textwidth]{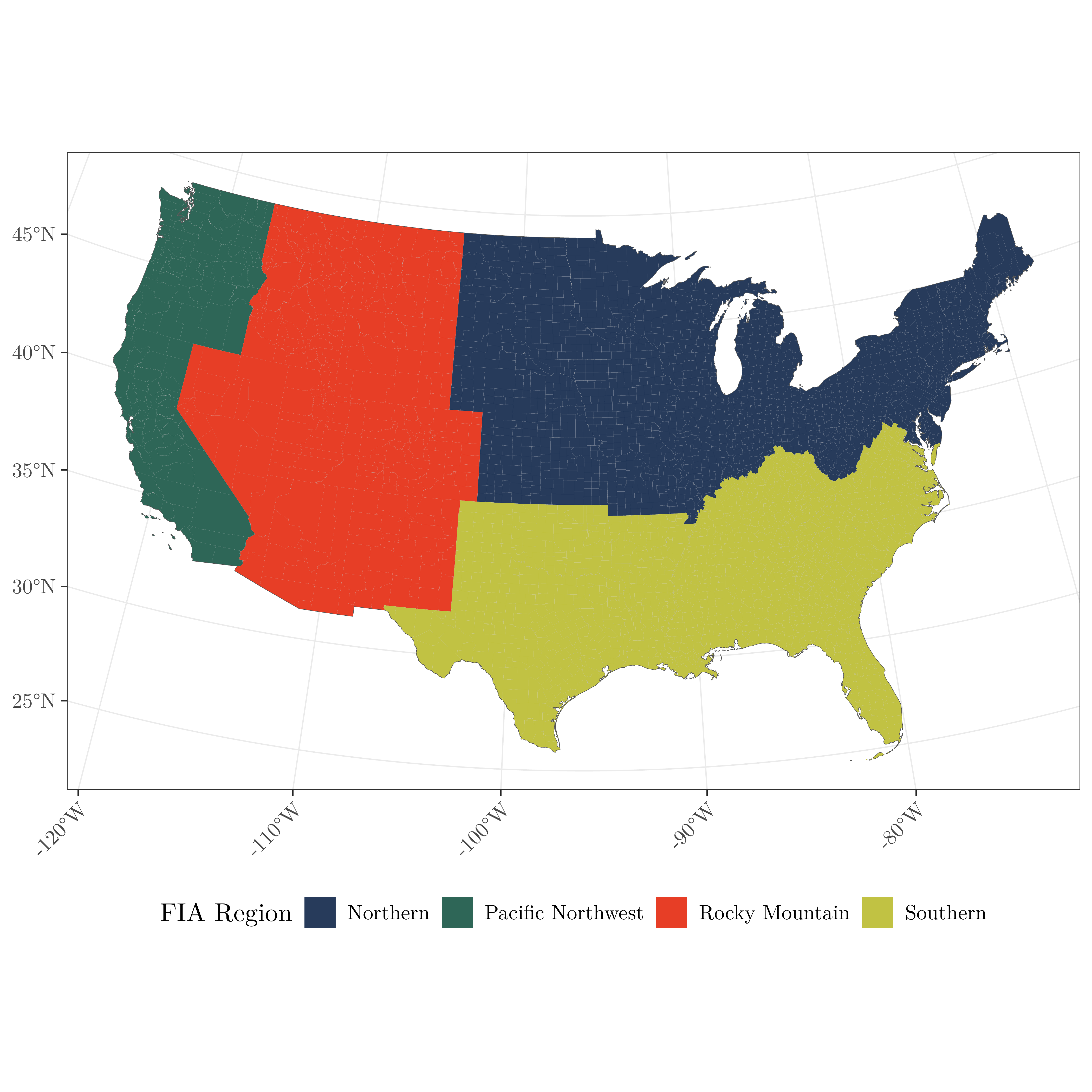}
\caption{FIA regions used to compute county aggregate total, trend, and change estimates given in Figure~\ref{fig:fia_regions_annual_totals} and Table~\ref{tab:fia_regions_real_change}.}\label{fig:fia_regions}
\end{figure}

\begin{table}[ht!]
\begin{center}
\begin{tabular}{lcc}
\toprule
  FIA Region & Trend $A_\mathcal{J}\theta_{\mathcal{J}}$ (Tg/year) &  Change $A_\mathcal{J}\Delta_{\mathcal{J}}$ (Tg)\\
  \cmidrule(lr){2-2} 
  \cmidrule(lr){3-3} 
  Northern          & -1.94 (-6.74, 3.42) & 43.03 (-43.05, 137.98)\\
  Pacific Northwest & -9.55 (-14.71, -5.34) & -101.01 (-186.91, -23.12)\\
  Rocky Mountain    & -12.86 (-19.46, -6.13) & -137.03 (-285.29, -1.29)\\
  Southern          & 44.32 (36.83, 50.55) & 660.45 (534.73, 781.08)\\
  \bottomrule
  \end{tabular}
  \caption{Left column, Full model estimates of total carbon trend by FIA region from 2008 and 2021, where $A_\mathcal{J}$ is the area of region $\mathcal{J}^{\text{th}}$ in hectares and $\theta_{\mathcal{J}}$ is carbon trend (Tg/ha/year) as defined in \eqref{eq:mu_aggregate}. Right column, Full model estimates of total carbon change by FIA region between $t_1$=2008 and $t_2$=2021, where $\Delta_{\mathcal{J}}$ is carbon change (Tg/ha). Estimates are posterior medians with 95\% credible intervals values given in parentheses. These trend and change estimates summarize patterns seen in Figure~\ref{fig:fia_regions_annual_totals}.}\label{tab:fia_regions_real_change}
  \end{center}
  \end{table} 

\clearpage
\subsection{County specific figures}\label{sec:county_supporting_figures}

Figures comparable to Figure~\ref{fig:real_ex_county_theta_pos_neg} for all counties are provided at \url{https://drive.google.com/drive/folders/1seksf153roVI6gFsiUKjr704Dk2Ru1xC?usp=sharing}. \emph{Upon acceptance, these figures will be moved to a permanent repository}.

\section{Data and code}\label{sec:supporting_code}

Data and code to reproduce results are provided at \url{https://drive.google.com/drive/folders/1seksf153roVI6gFsiUKjr704Dk2Ru1xC?usp=sharing}. \emph{Upon acceptance, these figures will be moved to a permanent repository}.

\section{MCMC sampler and computing notes}\label{sec:sampler_computing}

Here we provide the MCMC sampler for the full model defined in (\ref{mod:full}).

To ease notation, we move from the double indexing notation used in the main text to a single index notation. Specifically, we define the indexing set $i = 1, \ldots, N$, which directly maps to the double indexing set $(j, t)$ defined in Section (\ref{sec:direct}), whereby variables are ordered by time $t = 1, \ldots, T$ within county $j = 1, \ldots, J$. Following this new notation, let $\bmu = (\mu_1, \ldots,  \mu_N)^\top$ be the length $N$ vector of latent means and $\hat{\bmu} = (\hat{\mu}_1, \ldots,  \hat{\mu}_N)^\top$ be the corresponding length $N$ vector of direct estimates. Let $\bX$ be the $N \times (p+1)$ design matrix with rows corresponding to the ordering of $\bmu$ and columns corresponding to an intercept and $p$ many predictors $(\bone, \bx_1, \ldots, \bx_p)$, where $\bone$ is a column of ones and $\bx_k = (x_{k, 1}, \ldots,  x_{k, N})^\top$ for $k = 1, \ldots, p$. We then define the corresponding length $p + 1$ coefficient vector $\bbeta = (\beta_0, \ldots, \beta_p)^\top$. Let $\bet_0^{st} = (\eta^{st}_{0, 1}, \ldots, \eta^{st}_{0, N})^\top$ be the length $N$ vector corresponding to the spatial-temporal random intercept. Let $\bet^s$ be the length $Jq$ vector of random effects, with $\bet^s = (\eta_{1, 1}^s, \ldots \eta_{q, 1}^s, \ldots, \eta_{1, J}^s, \ldots, \eta_{q, J}^s)^\top$. To update the elements of $\bet^s$, we define $q$ many length $J$ sub-vectors, ($\bet_1^s, \ldots, \bet_q^s)$, where $\bet_k^{s} = (\eta_{k, 1}^{s}, \dots, \eta_{k, J}^{s})^\top$ for $k = 1, \ldots, q$.  We define the length $J(q-1)$ vector $\bet^s_{-k}$ to be $\bet^s$ with the elements of $\bet^s_k$ removed. To replicate the elements of $\bet^s$ over time, we define the $Nq \times Jq$ design (indicator) matrix $\bZ = \bI \otimes \bone$, where $\bI$ is a $Jq \times Jq$ identity matrix and $\bone$ is a length $T$ column vector of ones. Additionally, we define $\bZ_k = \bI \otimes \bone$ where $\bI$ is a $J \times J$ identity matrix and $\bone$ is a length $T$ column vector of ones and $\bZ_{-k} = \bI \otimes \bone$ where $\bI$ is a $J(q-1) \times J(q-1)$ identity matrix and $\bone$ is a length $T$ column vector of ones. Finally, we let $\tilde{\bX}$ be the $N \times Nq$ block diagonal design matrix of spatially-varying predictors. Specifically, the $i$-th block is the length $q$ vector $(\tilde{x}_{1, i}, \ldots, \tilde{x}_{q, i})$.   Accordingly, we define $\tilde{\bX}_k$ to be the $N \times N$ diagonal matrix with elements $(\tilde{x}_{k, 1}, \ldots, \tilde{x}_{k, N})$ for $k = 1, \ldots, q$, and $\tilde{\bX}_{-k}$ to be the $N \times N(q-1)$ block diagonal matrix with blocks equal to those in $\tilde{\bX}$ with the $k^\text{th}$ element removed. The full model in (\ref{mod:full}) may then be written as 
\begin{equation}
    \bmu \sim MVN(\bet_0^{st} + \bX\bbeta + \tilde{\bX}\bZ\bet^s, \sigma_\epsilon\bI)
\end{equation}
and update parameters as follows.
\begin{enumerate}
    \item Update elements of $\bmu$, where corresponding $\hat{\mu}_i$ and $\hat{\sigma}^2_i$ (hence $\sigma^2_{i}$) are available, using $\bmu\given\cdot \sim MVN(\bV\bv, \bV)$ where $\bv = \bSigma_\mu^{-1}\hat{\bmu} + (\bet_0^{st} + \bX\bbeta + \tilde{\bX}\bZ\bet^s)/\sigma^2_\epsilon$, $\bV^{-1} = (\bSigma^{-1}_\mu + \sigma^{-2}_\epsilon\bI)$, and $\bSigma_\mu$ is diagonal with diagonal elements $\sigma^2_{i}$s. Vector and matrix dimensions are adjusted accordingly to remove rows corresponding to missing values. Note, $\bV$ is diagonal so, for efficiency, sampling should be done using a univariate normal.

    \item Update elements of $\bmu$, where either corresponding $\hat{\mu}_{i}$ or $\hat{\sigma}^2_{i}$ is \textbf{not} available, using $\bmu\given\cdot \sim MVN(\bet_0^{st} + \bX\bbeta + \tilde{\bX}\bZ\bet^s, \sigma^2_\epsilon\bI)$. Vector and matrix dimensions are adjusted accordingly to include rows corresponding to missing values. Note, the variance is diagonal so, for efficiency, sampling should be done using a univariate normal.

    \item Update $\bbeta$ using $\bbeta\given\cdot \sim MVN(\bV\bv, \bV)$ where $\bv = \left(\bSigma^{-1}_\beta\bmu_\beta + \bX^\top(\bmu - \bet_0^{st} - \tilde{\bX}\bZ\bet^s)/\sigma^2_\epsilon\right)$, $\bV^{-1} = \left(\bSigma^{-1}_\beta + \bX^\top\bX/\sigma^2_\epsilon\right)$ and $\bSigma_{\beta} = \sigma^2_{\beta}\bI$.

    \item Update $\bet^{st}_0$ using $\bet^{st}_0\given\cdot \sim MVN(\bV\bv, \bV)$ where $\bv = (\bmu - \bX\bbeta - \tilde{\bX}\bZ\bet^s)/\sigma^2_\epsilon$ and $\bV^{-1} = \sigma^{-2}_{\eta^{st}_0}\bR(\rho_{\eta^{st}_0})^{-1}\otimes\bA(\alpha_{\eta^{st}_0})^{-1} + \sigma^{-2}_\epsilon\bI$. 

    \item Update $\bet^{s}_k$ for $k = 1, \ldots, q$ using $\bet^{s}_k\given\cdot \sim MVN(\bV\bv, \bV)$ where $\bv = (\tilde{\bX}_{k}\bZ_k)^\top(\bmu - \bet_0^{st} - \bX\bbeta - \tilde{\bX}_{-k}\bZ_{-k}\bet^{s}_{-k})/\sigma^2_\epsilon$ and $\bV^{-1} = \sigma^{-2}_{\eta^s_k}\bR(\rho_{\eta^s_k})^{-1} + \sigma^{-2}_\epsilon(\tilde{\bX}_k \bZ_k)^\top(\tilde{\bX}_k \bZ_k)$.

    \item Update $\sigma^2_i$ for $i$ where both $\hat{\mu}_i$ and $\hat{\sigma}^2_i$ are available using $\sigma^2_i\given\cdot \sim IG(a, b)$ where $a = n_i/2 + 1/2$ and $b = (n_i - 1)\hat{\sigma}_i^2/2 + (\hat{\mu}_i - \mu_i)^2/2$.

    \item Update $\sigma^2_{\eta^{st}_0}$ using $\sigma^2_{\eta^{st}_0}\given\cdot \sim IG(a, b)$ where $a = a_{\eta^{st}_0} + N/2$ and $b = b_{\eta^{st}_0} + (\bet^{{st}\top}_0(\bR(\rho_{\eta^{st}_0})\otimes\bA(\alpha_{\eta^{st}_0}))^{-1}\bet^{st}_0)/2$.

    \item Update $\sigma^2_{\eta^s_k}$ for $k = 1,\ldots,q$ using $\sigma^2_{\eta^s_k}\given\cdot \sim IG(a, b)$ where $a = a_{\eta^s} + J/2$ and $b = b_{\eta^s} + \bet^{{s}\top}_k\bR(\rho_{\eta^s_k})^{-1}\bet^{s}_k$.

    \item Update $\sigma^2_\epsilon$ using $\sigma^2_\epsilon\given\cdot \sim IG(a, b)$ where $a = a_\epsilon + N/2$ and $b = b_\epsilon + (\bet^{st}_0 + \bX\bbeta + \tilde{\bX}\bZ\bet^{s})^\top(\bet^{st}_0 + \bX\bbeta + \tilde{\bX}\bZ\bet^{s})/2$.

    \item Jointly update $\rho_{\eta^{st}_0}$ and $\alpha_{\eta^{st}_0}$ using the Metropolis algorithm with log target density plus Jacobian adjustment for the Uniform prior distribution proportional to
    \begin{multline*}
        -1/2\log|\sigma^2_{\eta^{st}_0}\bR(\rho_{\eta^{st}_0})\otimes\bA(\alpha_{\eta^{st}_0})|-1/2\bet^{st\top}_0(\sigma^2_{\eta^{st}_0}\bR(\rho_{\eta^{st}_0})\otimes\bA(\alpha_{\eta^{st}_0}))^{-1}\bet^{st}_0 +\\
        \log(\alpha_{\eta^{st}_0} - a_\alpha) + \log(b_\alpha - \alpha_{\eta^{st}_0}) +
        \log(\rho_{\eta^{st}_0} - a_\rho) + \log(b_\rho - \rho_{\eta^{st}_0}).
    \end{multline*}

    \item Update $\rho_{\eta^s_k}$ for $k = 1, \ldots, q$ using the Metropolis algorithm with log target density plus Jacobian adjustment for the Uniform prior distribution proportional to
    \begin{multline*}-1/2\log|\sigma^2_{\eta^{s}_k}\bR(\rho_{\eta^{s}_k})|-1/2\bet^{s\top}_k(\sigma^2_{\eta^{s}_k}\bR(\rho_{\eta^{st}_0}))^{-1}\bet^{s}_k +
        \log(\rho_{\eta^{s}_k} - a_\rho) + \log(b_\rho - \rho_{\eta^{s}_k}).
    \end{multline*}
      
\end{enumerate}

To efficiently evaluate the CAR precision matrix $\bR(\rho_{\eta^s})^{-1}$ defined in Section~\ref{sec:model}, we define $\bR(\rho_{\eta^s})^{-1} = (\bD - \rho_{\eta^s} \bW) = \bD^{1/2}(\bI - \rho_{\eta^s} \bD^{-1/2}\bW \bD^{-1/2}) \bD^{1/2}$, where $\bI$ is the $J \times J$ identity matrix, and let $\bD^{-1/2} \bW \bD^{-1/2} = \bP\bLambda \bP^{\top}$ where $\bLambda$ is the diagonal matrix of eigenvalues and the columns of $\bP$ are the eigenvectors of $\bD^{-1/2} \bW \bD^{-1/2}$. This allows $\bR(\rho_{\eta^s})^{-1}$ to be expressed as $\sum_{i=1}^J (1 - \rho_{\eta^s}\lambda_i)\bv_i \bv_i^{\top} = \sum_{i=1}^J \bv_i \bv_i^{\top} - \rho_{\eta^s} (\sum_{i=1}^J\lambda_i \bv_i \bv_i^{\top})$, where $\bv_i$ are the columns of $\bD^{1/2}\bP$ and $\lambda_i$ are the diagonal elements of $\bLambda$. Expressing the precision matrix in this way removes the need for costly matrix formation and Cholesky decomposition in each MCMC iteration. Rather, the first term, i.e., $\sum_{i=1}^J \bv_i \bv_i^{\top}$, remains the same across MCMC iterations and the second term, i.e., $\rho_{\eta^s} (\sum_{i=1}^J\lambda_i \bv_i \bv_i^{\top})$, only varies by a multiplicative constant $\rho_{\eta^s}$. Further, the determinant of $\bR(\rho_{\eta^s})^{-1}$, which is needed to update correlation parameters $\rho_{\eta^s}$ and $\rho_{\eta_0^{st}}$ via Metropolis steps, is simplified to $\prod^J_{i=1}\bD_{ii}\left(1 - \rho_{\eta^s}\lambda_i\right)$, which requires no linear algebra across MCMC iterations. 

In Steps 3, 4, and 5 we sample from $MVN(\bV\bv, \bV)$ which is most efficiently accomplished via $\bU^{-1}\bb$ where $\bb = \bU^{-\top}\bv + \ba$, $\bU$ is the upper-triangular Cholesky square root of $\bV^{-1}$ and $\ba$ is a vector of realizations from a standard normal distribution $N(0,1)$. 

In step 10, we utilize matrix sparseness and properties of Kronecker products and determinants to compute $\log(|\sigma^2_{\eta^{st}_0}\bR(\rho_{\eta^{st}_0})\otimes\bA(\alpha_{\eta^{st}_0})|)$. Let $\bA(\alpha_{\eta^{st}_0}) = \bB \bB^\top$ be the Cholesky decomposition of $\bA(\alpha_{\eta^{st}_0})$, with $\bb$ as the vector of diagonal elements of $\bB$. Also, let $\bd$ and $\blambda$ be the vectors of diagonal elements of the matrices $\bD$ and $\bLambda$ defined previously. We then have
\begin{align*}
\log|\sigma^2_{\eta^{st}_0}\bR(\rho_{\eta^{st}_0})\otimes\bA(\alpha_{\eta^{st}_0})| & = \log(|\bA(\alpha_{\eta^{st}_0})|^J \times |\sigma^2_{\eta^{st}_0}\bR(\rho_{\eta^{st}_0})|^T) \\
& = J \log|\bA(\alpha_{\eta^{st}_0})| + T \log |\sigma^2_{\eta^{st}_0}\bR(\rho_{\eta^{st}_0})| \\
& = 2J \sum \log\bb -T \log |1/\sigma^2_{\eta^{st}_0}\bR(\rho_{\eta^{st}_0})^{-1}| \\
& = 2J \sum \log\bb -T (J \log (1/\sigma^2_{\eta^{st}_0}) + \log |\bR(\rho_{\eta^{st}_0})^{-1}|) \\
& = 2J \sum \log\bb - N \log (1/\sigma^2_{\eta^{st}_0}) - T \sum \log (\bd \odot (1 - \rho_{\eta_0^{st}} \blambda)), 
\end{align*}
where $\odot$ denotes element-wise multiplication. 

In step 11, we again use the same identities to efficiently evaluate $\log|\sigma^2_{\eta^{s}_k}\bR(\rho_{\eta^{s}_k})|$ as 
\begin{align*}
\log|\sigma^2_{\eta^{s}_k}\bR(\rho_{\eta^{s}_k})| & = - \log |1 / \sigma^2_{\eta^{s}_k}\bR(\rho_{\eta^{s}_k})^{-1}| \\ & = - J \log (1/\sigma^2_{\eta^{s}_k}) - \sum \log(\bd \odot (1 - \rho_{\eta^{s}_k} \blambda)).
\end{align*}

\section*{Acknowledgements}

This work was supported by the USDA Forest Service, NSF DEB-2213565, and NASA CMS grants Hayes (CMS 2020). The findings and conclusions in this publication are those of the author(s) and should not be construed to represent any official US Department of Agriculture or US Government determination or policy.

\bibliographystyle{apalike}
\bibliography{literature}

\end{document}